\documentclass[prb,preprint]{revtex4-1}
\usepackage[T1]{fontenc}
\usepackage{graphicx}
\usepackage{amssymb}
\usepackage{amsmath}
\usepackage{epstopdf}
\usepackage{xfrac}

\DeclareMathAlphabet{\mathbbold}{U}{bbold}{m}{n}
\usepackage{subfigure}
\DeclareRobustCommand*{\citen}[1]{%
  \begingroup
    \romannumeral-`\x 
    \setcitestyle{numbers}%
    \cite{#1}%
  \endgroup   
}

\newcommand{\figpath}{./}
\newlength{\figwidth}
\newcommand{\aref}[1]{App.\,\ref{#1}}
\newcommand{\fref}[1]{Fig.\,\ref{#1}}
\newcommand{\tref}[1]{Table\,\ref{#1}}
\newcommand{\eref}[1]{Eq.\,(\ref{#1})}

\newcommand{\sref}[1]{Sec.\!~\ref{#1}}

\newcommand{\cref}[1]{Ref.\,\citen{#1}}
\newcommand{\crefs}[1]{Refs.\,\citen{#1}}
\newcommand{\cf}{{\it cf.}\ } 
\newcommand{\vs}{{\it vs.}\ }

\newcommand{\eg}{{\it e.g.}\,}

\newcommand{\etal}{{\it et al.}\ }

\newcommand{\abinitio}{{\it ab initio} }

\usepackage{color}

\newcommand{\Vr}{\mathrm{V}}
\newcommand{\Sb}{\mathbf{S}}

\newcommand{\eb}{\mathbf{e}}
\newcommand{\ub}{\mathbf{u}}
\renewcommand{\Re}{\mathrm{Re}}
\newcommand{\dm}{{\mathrm{d}}}

\newcommand{\tcoef}{\eta}
\newcommand{\virial}{{\boldsymbol{\nu}}}

\newcommand{\heatfluxb}{{\mathbf{J}}}

\newcommand{\heatfluxv}{{\mathbf{J}}}

\newcommand{\conductivityb}{{\boldsymbol{\kappa}}}

\newcommand{\Fb}{\mathbf{F}}
\newcommand{\ab}{\mathbf{a}}

\newcommand{\fb}{\mathbf{f}}

\newcommand{\vb}{\mathbf{v}}

\newcommand{\xb}{\mathbf{x}}
\newcommand{\kb}{\mathbf{k}}
\newcommand{\yb}{\mathbf{y}}
\newcommand{\pb}{\mathbf{p}}
\newcommand{\nb}{\mathbf{n}}
\newcommand{\Ib}{\mathbf{I}}

\newcommand{\Jb}{\mathbf{J}}

\newcommand{\Eb}{\mathbf{E}}
\newcommand{\Pb}{\mathbf{P}}
\newcommand{\Mb}{\mathbf{M}}
\newcommand{\Xb}{\mathbf{X}}

\newcommand{\Ac}{\mathcal{A}}

\newcommand{\epsilonb}{{\boldsymbol{\epsilon}}}
\newcommand{\sigmab}{\boldsymbol{\sigma}}

\newcommand{\grad}{\boldsymbol{\nabla}}

\newcommand{\erf}{\operatorname{erf}}
\newcommand{\erfc}{\operatorname{erfc}}

\newcommand{\Fc}{\mathcal{F}}
\renewcommand{\Bbb}{\mathbb{B}}
\newcommand{\cbb}{{\mathbbold{c}}}
\newcommand{\bbb}{{\mathbbold{b}}}
\newcommand{\Cbb}{\mathbb{C}}
\newcommand{\Dbb}{\mathbb{D}}
\newcommand{\Kbb}{\mathbb{K}}
\newcommand{\tr}{\operatorname{tr}}

\begin{document}
\title{\bf Estimates of crystalline LiF thermal conductivity at high temperature and pressure by a Green-Kubo method}
\author{R.E. Jones }
\affiliation{\it Sandia National Laboratories, P.O. Box 969, Livermore, CA 94551, USA}
\email{corresponding author: rjones@sandia.gov} 
\author{D.K. Ward}
\affiliation{\it Sandia National Laboratories, P.O. Box 969, Livermore, CA 94551, USA}

\begin{abstract}
Given the  unique optical properties of LiF, it is often used as an observation window in high-temperature and pressure experiments; and, hence, estimates of its transmission properties are necessary to interpret observations.
Since direct measurements of the thermal conductivity of LiF at the appropriate conditions are difficult, we resort to molecular simulation methods.
Using an empirical potential validated against \abinitio phonon density of states,  we estimate the thermal conductivity of LiF at high temperatures (1000--4000K) and pressures (100--400 GPa) with the Green-Kubo method. 
We also compare these estimates to those derived directly from \abinitio data.
To ascertain the correct phase of LiF at these extreme conditions we calculate the (relative) phase stability of the B1 and B2 structures using a quasiharmonic \abinitio model of the free energy.
We also estimate the thermal conductivity of LiF in an uniaxial loading state that emulates  initial stages of compression in high-stress ramp loading experiments and show the degree of anisotropy induced in the conductivity due to deformation.
\end{abstract}

\maketitle
\section{Introduction}

LiF is a ionic solid that is particularly transparent to short wavelength radiation due to its large band gap and hence is commonly used in optics for high-pressure and temperature experiments, such as those related to the development of pulsed power \cite{z-machine}.
LiF is also a component in molten salts frequently employed as high-temperature thermal fluids.
Estimates of the transport properties of LiF are important to both these applications. 
Specifically, in dynamic high-pressure experiments, a LiF window maintains the pressure at the sample interface where velocimetry measurements are typically made.  
Due to the extreme conditions, the necessary transmission properties are difficult to measure directly.
In these experiments, a shock or a near-shock ramp compression with pressures up to 800 GPa \cite{fratanduono2011refractive} is generated by a variety of means and the material response is measured  using velocity interferometry, see, \eg, \crefs{rigg1998real,knudson2004principal,lalone2008velocity,hicks2009laser,fratanduono2011refractive}.
There are many efforts concentrating on estimating the optical properties of LiF crucial to this measurement, see, \eg, \crefs{jensen2007accuracy,fratanduono2011refractive,rigg2014determining,spataru2015ab}.
Due to the short but finite timescale of dynamic material experiments, the thermal conductivity of LiF windows can significantly affect the temperature measured at the sample interface.  In this work, we focus on calculating this thermal conductivity at the extreme conditions relevant to these experiments with the goal of understanding the nonequilibrium energy transfer that governs their behavior.

The material properties of LiF have been explored in experiments and simulation primarily nearer to ambient conditions.
For instance, Thacher \cite{thacher1967effect} measured the sound velocity and thermal conductivity of LiF at temperatures less than 100 K. 
At ambient pressure, \cref{thacher1967effect} shows that the  thermal conductivity of LiF peaks at about 2 W/m-K near 20 K, where the quantum increase of heat capacity begins to be dominated by the decrease in conductivity due to Umklapp scattering.
Andersson and B\"ackstr\"om \cite{andersson1987thermal} were able to measure the heat capacity and thermal conductivity of LiF at room temperature up to pressures of 1 GPa. 
They showed a linear dependence of thermal conductivity on pressure and measured a conductivity value of 16.3 W/mK at 1 GPa.
Phase and other transitions can complicate measurements at higher temperatures and pressures. 
Given that the full phase diagram for LiF is not known, Smirnov \cite{Smirnov2011abinitio} computed an \abinitio phase diagram of LiF over pressures ranging from 0 to 500 GPa and temperatures ranging from 0 to 12000 K together with the elastic properties and associated Debye temperatures.
Smirov  \cite{Smirnov2011abinitio} predicted that the structure of LiF is the NaCl-like arrangement B1 at low temperature and pressure but transitions to the CsCl-like arrangement B2 at higher pressures and temperatures.
Smirov correlated his results with experimental data by Kormer \cite{kormer1968optical} and Boehler \etal \cite{boehler1997melting}.
(See Root \etal \cite{root2015shock} for a similar study of MgO where \abinitio molecular dynamics and quantum Monte Carlo methods were also employed to predict a phase diagram at extreme conditions.)
Boehler \etal \cite{boehler1997melting} studied the high pressure melting regime of LiF with diamond anvil experiments and classical molecular dynamics (MD) fitted to properties at standard temperature and pressure.
Cl\'erouin \etal \cref{clerouin2005ab} used \abinitio dynamics to estimate the  optical properties of LiF along the shock Hugoniot where it transitions from a transparent solid to a reflective plasma.
Belonoshko \etal \cite{Belonoshko2000Born} investigated LiF melting with MD using a potential tuned with \abinitio data and compare to existing diamond anvil and shock experimental data.
In particular, Belonoshko \etal showed that density as a function of pressure and the radial distribution function computed with their potential compares well with trusted data.
They also make clear the distinction between thermal instability and melting especially for small systems at high pressures using a phase coexistence method.
Also using classical MD, Young \cite{Young2004Buck} studied ion damage of LiF crystals.
Related to thermal properties of LiF,
N\"{u}sslein and Schr\"{o}der \cite{nusslein1967calculations} calculated the dispersion and phonon density of states (phDOS) via polarizable model of the inter-atomic interactions of LiF at 0 K.
Dolling \etal \cite{dolling1968lattice} also calculated the phDOS of LiF via lattice dynamics and compared it to dispersion data derived from slow neutron inelastic scattering.
In their work, the crystal has phonon content up to 20 THz with most of the low frequency content attributed to the F ion.
Recently, Stegailov \cite{stegailov2010stability} calculated the phDOS with density functional theory with the generalized gradient approximation and showed the onset of mechanical instability, which may lead to defect formation or melting, due to hot electrons when the electron temperature reaches 37,000 K.

Following this body of work, in this paper we use MD together with the Green-Kubo (GK) formalism \cite{Onsager1931a,Onsager1931b, Green1954, Kubo1957, Kubo1957b, Zwanzig1964} to estimate the thermal conductivity of LiF 
at stresses on par with the elastic moduli and temperatures in excess of the melt temperature at ambient conditions.
In particular, we investigate both volumetric and uniaxial deformation modes similar to (non-Hugniot) ramp compression experiments.
(\cref{spataru2015ab} makes a corresponding study of the optical properties of LiF.)
Generally speaking, classical MD is well-suited to simulating ionic solids since essentially all heat propagates via phonons. 
The related publications, \crefs{ciccotti1976transport,Lindan1991conductivity,Galamba2004conductivity,Ishii2014conductivity,ohtori2009calculations,salanne2011thermal,ishii2014thermal}, are the few examples of estimating the thermal conductivity of alkalis with classical MD and typically focus on the thermal properties of molten salts with applications to high temperature thermal transfer fluids.
The strong ionic character of LiF leads to the usual complications due to long range Coulomb interaction, requiring dipole corrections and large cell sizes; but the main issue is that MD is highly reliant on empirical potentials.
There has been some work on suitable potentials for ionic solids like LiF, typically of the Buckingham \cite{Young2004Buck,Cherednikov2013defects} or Born/Tosi-Fumi \cite{Tosi1963revise,tosi1964ionic,jacucci1976effects} forms.
A high quality potential parameterization for LiF of another form was developed by Ishii \etal \cite{Ishii2014conductivity} but focussed on the properties of molten mixtures, see also \crefs{ohtori2009calculations,salanne2011thermal,ishii2014thermal}.
On the other-hand,  Belonoshko \etal \cite{Belonoshko2000Born} carefully constructed a Tosi-Fumi/Born-Mayer-Huggins potential to suit high-pressure and temperature conditions by dropping the unstable terms in the full Tosi-Fumi form, and  compared its behavior to literature and their own density functional theory (DFT) results.
Given the findings in \cref{Belonoshko2000Born}, it was evident that the MD potential may not transition to the most stable phase with changes in pressure and temperature but instead becomes stuck in a metastable state. 
We used this fact together with a phase diagram independently calculated with DFT to estimate thermal conductivity over pressures ranging from 100 to 400 GPa and temperatures ranging from 1000 to 4000 K.
To compute the phase diagram, we follow Smirnov's work \cite{Smirnov2011abinitio} and others \cite{correa2008first,correa2006carbon,knudson2008shock} and use plane augmented wave (PAW) DFT with the local density approximation (LDA) instead of the linear muffin tin orbital method Smirnov employed to estimate the zero temperature enthalpy and entropy of the phonon population.
From a dynamical matrix calculated with DFT, we are able to estimate the entropy component of the free energy with a quasi-harmonic model.
The range of the free energy estimated with the quasi-harmonic model  limited by the  mechanical stability which we also estimated with the \abinitio bond stiffnesses governing the phonon propagation.  
In addition, we use the \abinitio phDOS to validate and recalibrate the Belonoshko parameterization for thermal conductivity estimates.

\section{Theory} \label{sec:theory}
Given a definition of the heat flux $\Jb$, the thermal conductivity tensor $\conductivityb$ can be obtained from the Green-Kubo formula dependent on the time-correlation of $\Jb$ with itself:
\begin{equation}\label{eq:GK_conductivity}
\conductivityb \ = \ \dfrac{V}{k_B T^2} \int_0^{\infty} \Bigl< \heatfluxb(0) \otimes \heatfluxb(t) \Bigr> \, \dm t  \ ,
\end{equation}
where $V$ is the system volume, $T$ is the temperature, $k_B$ is the Boltzmann constant. 
The bracket $\langle \cdot \rangle$ denotes the appropriate ensemble average, where it is important to note that $\langle \heatfluxv \rangle = \mathbf{0}$ in equilibrium.
A microscopic formula \cite{Irving1950,Mandadapu2009} for the heat flux $\Jb$ is
\begin{equation}
\heatfluxb = \frac{1}{V} \sum_\alpha \left( \varepsilon_\alpha \Ib + \boldsymbol{\nu}_\alpha^T \right) \vb_\alpha \ ,
\end{equation}
where the per-atom energy $\epsilon_\alpha$ is formed from the kinetic energy of the atom $\alpha$ and a reasonable partition of the total potential energy comprised of short-range bonds and long-range Coulomb interactions to individual atoms \cite{Schelling2002}, and the virial stress $\virial_\alpha$ for atom $\alpha$ in terms of fundamental quantities (which is given in \aref{app:virial}).
Classical molecular dynamics (MD) provides the necessary positions $\xb_\alpha$, velocities $\vb_\alpha$, and forces $\fb_\alpha$, from the trajectories obtained by integrating Newton's equations of motion, $m_\alpha \ddot{\xb}_\alpha = \fb_\alpha$, given an initial configuration $\{\xb_\alpha(0)\}$ and atomic masses $m_\alpha$. 
The total force $\fb_\alpha $ on an atom $\alpha$ is the sum of interatomic forces derived from an empirical potential $\Phi ( \{ \xb_\alpha \} )$.
For ionic solids like LiF, explicit charges $q_\alpha$ are typically constant and located at ion cores.

A widely-used potential for ionic materials is the Tosi-Fumi/Born-Mayer-Huggins (TF/BMH) potential \cite{tosi1964ionic},\cite{Huggins1933M1,Huggins1937M2,Huggins1947M3}.
It is a combination of long-range Coulomb and short-range $\varphi$ (repulsion) pair-wise interactions
\begin{equation} \label{eq:Born}
\Phi = \sum_{a\le b} \sum\limits_{\substack{\alpha \in \Ac_a\\ \beta \in \Ac_b}}^{\alpha\neq \beta} \phi_{ab}(r_{\alpha\beta}) \ \ \ \text{where} \ \ \
\phi_{ab}(r)=
\frac{z_a z_b e^2}{\epsilon r}
+\underbrace{A_{ab}\exp(-B_{ab}r)
-\frac{C_{ab}}{r^{6}}-\frac{D_{ab}}{r^{8}}}_{\varphi_{ab}(r)}
\end{equation}
for species $a$,$b$ with associated groups of atoms $\Ac_a$,$\Ac_b$; inter-atomic distance $r_{\alpha\beta} = \| \xb_\alpha - \xb_\beta \|$; and charge $q_\alpha = z_a e$ for $\alpha \in \Ac_a$.
Here, $\epsilon$ is the (vacuum) permittivity and $e$ is the elementary charge.
Of the empirical parameters: $A_{ab}$, $B_{ab}$, $C_{ab}$ and $D_{ab}$, the last two are related to dipole interactions and are set to zero for high pressure stability considerations by Belonoshko \etal \cite{Belonoshko2000Born} in their model of LiF.
The periodic images participating in the Coulomb forces on the atoms in the system extend well beyond the explicitly represented periodic box.
For efficiency, the energy $\Phi$ is decomposed into long (\emph{reciprocal} $\kb$-space) and short (real $\xb$-space) components.
This decomposition is the essence of Ewald summation and the Particle-Particle Particle-Mesh (PPPM) method \cite{ewald1921berechnung,hockney2010computer,karasawa1989acceleration,heyes1994pressure,sirk2013characteristics} (see \aref{app:virial} for more details).

To validate an empirical potential for calculation of thermal conductivity ideally all the properties related to the phonon population and propagation would be compared with experimental or \abinitio data. 
In lieu of a full comparison of the dispersion relationship for harmonic waves and related properties for anaharmonic interactions, we follow others in comparing the elastic constants and phonon density of states related to the phonon dispersion and wave speeds.
As derived from the dispersion relationship of the material, the phonon density of states (phDOS) is linked to the thermal conductivity of the material.
The dispersion relationship is determined by the matrix of bond stiffnesses $\Kbb$, which is composed of sub-matrices of linearized force constants:
\begin{equation} \label{eq:Kmat}
\left[\Kbb\right]_{\alpha\beta} 
= \left.  \frac{\partial^2 \Phi}{\partial \xb_\alpha \partial \xb_\beta} \right|_{\xb_\alpha=\yb_\alpha}  
= - \left. \frac{\partial \fb_\alpha}{\partial \xb_\beta} \right|_{\xb_\alpha=\yb_\alpha}
\end{equation}
referenced to a given lattice configuration $\yb_\alpha$ \cite{giannozzi1991ab,kresse1995ab,gonze1997dynamical}.
The dynamical matrix, a Fourier transform of $\Kbb$, results from applying a plane  wave ansatz for the motion of the atoms about lattice positions $\yb_\alpha$:
\begin{equation} \label{eq:Dmat}
\left[\Dbb(\kb)\right]_{\alpha\beta} 
= \sum_{\boldsymbol{\ell}} \frac{1}{\sqrt{m_\alpha m_\beta}} \Kbb_{\alpha\beta}
\exp\left(-\imath \kb \cdot \left( \xb_{\alpha} - \xb_{\beta} - \boldsymbol{\ell} \right) \right) \ ,
\end{equation}
where $\boldsymbol{\ell}$ ranges across all periodic images of the unit cell including the original one.
The dynamical matrix determines the eigenvalues $\omega_i^2$ for a given propagation direction (wave vector) $\kb$ and polarization.
The phDOS is constructed by sampling the eigenvalues $\omega_i^2$ of \eref{eq:Dmat} throughout the Brillouin zone.
The same procedure can be used in the context of an \abinitio density functional model of the material where the forces $\fb_\alpha$ are the Hellman-Feynman forces.
The dynamical matrix also determines the (linear) phonon and long-wavelength elastic stability.
The elastic moduli tensor
\begin{equation}
\left[ \Bbb \right]_{iAjB} 
=  \left[\frac{\partial^2 \Phi}{\partial {\Fb} \partial {\Fb} } \right]_{iAjB}
= \frac{1}{\Vr_0} \left[ \sum_{\alpha,\beta} \frac{\partial^2 \Phi} {\partial {\xb_\alpha} \partial {\xb_\beta}} :
\frac{\partial \xb_\alpha}{\partial \Fb} \frac{\partial \xb_\beta}{\partial \Fb}
\right]_{iAjB}
= \frac{1}{\Vr_0} \sum_{\alpha,\beta} \left[ \left[ \Kbb \right]_{\alpha \beta} \right]_{ij}
\left[ \Xb_{\alpha} \right]_A  \left[ \Xb_{\beta} \right]_B
\end{equation}
is related to the tensor of bond stiffnesses $\Kbb$ and determines the stability in the continuum limit.
Here, $\Fb = \frac{\partial \xb}{\partial \Xb}$ is the deformation gradient, $\Xb_\alpha$ are the stress-free lattice sites, and $\yb_\alpha = \Fb \Xb_\alpha$.
See \aref{app:stability} for more details.

LiF can change phase over a range of temperatures and pressures.
To determine the relatively stable phase as a function of temperature $T$ and stress $\Pb$, estimates of the Gibbs free energy $G$:
\begin{equation}
G(\Pb,T) = F(\Fb(\Pb,T),T) + \Pb\cdot\Fb(\Pb,T) \ , 
\end{equation}
a Legendre transform of Helmholtz free energy $F$,
needs to be calculated for both B1 and B2 structures.
Here, $\Pb = \left. \partial_\Fb F \right|_T$ is the first Piola-Kirchhoff stress.
Assuming positive frequencies $\omega_i$,
the Helmholtz free energy $F$ is commonly estimated with a quasi-harmonic (QH) model:
\begin{equation} \label{eq:QH}
F(\Fb,T) 
= {E_c(\Fb) + \frac{1}{2} \sum \hbar \omega_i(\Fb)}
+ k_B T \sum_i \ln\left( 1- \exp\left(\frac{\hbar \omega_i(\Fb)}{k_B T} \right) \right) \ ,
\end{equation}
which is based on the partition function of independent oscillators (see \eg \cref{dove1993introduction}(Chap. 5)).
This model is composed of two zero temperature components: (a) the cohesive energy (referenced to an infinitely dispersed state) which can be equated with $E_c(\Fb) = \Phi(\{ \Fb\Xb_\alpha \})$, and (b) the (non-classical) zero point/ground state energy of the phonons $ \frac{1}{2} \sum \hbar \omega_i(\Fb) = \frac{1}{2} \hbar \tr \sqrt{\Dbb}$; together with a third term: the temperature-entropy product approximated by the harmonic oscillator model.
Given the wide band-gap of LiF, we neglect the thermal electron contribution in this approximate model.
Clearly, an equation of state (EOS) $\Pb = \Pb(\Fb,T)$ is necessary to transform the Helmholtz free energy $F$ to the Gibbs free energy $G$.
At zero temperature the data needed to construct an accurate EOS can be calculated with DFT.
The change of stress with temperature can be estimated with the QH model \eref{eq:QH} \cite{kimmer2007continuum} or from MD simulations.
Assuming a first order dependence of stress on temperature, we can use
\begin{equation} \label{eq:thermal_expansion}
\Pb = \Pb_0(\Fb) + \Mb T
\end{equation}
to form the necessary inverse $\Fb = \Pb_0^{-1}(\Pb-\Mb T)$.
Since the systems of interest are cubic, the thermal expansion tensor $\Mb \equiv \frac{\partial^2 F}{\partial T \partial \Fb} = \tcoef \Ib$, and hence only one coefficient $\tcoef$ needs to be determined to effect thermal expansion.

\section{Methods} \label{sec:method}

As discussed in the introduction, we have based this study on the potential by Belonoshko \etal \cite{Belonoshko2000Born}.
This potential was specifically parameterized for high pressure states where the $C_{ab}$ and $D_{ab}$ parameters of the TF/BMH potential in \eref{eq:Born}, which cause instability, are set to zero.
Since Belonoshko \etal \cite{Belonoshko2000Born} were primarily concerned with investigating phase diagram and mechanical properties and we are employing the potential to estimate thermal conductivity, we compared the phDOS resulting from the TF/BMH potential with the Belonoshko \etal parameters to that from an \abinitio calculation as a measure of the validity of phonon transmission.
For the DFT calculations, we employed the local density approximation (LDA), a plane augmented wave basis with cutoff 800 eV with standard pseudo-potentials \cite{kresse1999ultrasoft}, and 20$\times$20$\times$20 k-point Monkhorst-Pack grid, which were arrived at via convergence studies for the dynamical matrix and elastic moduli.

\begin{figure}[h!]
\centering
\subfigure[B1]
{\includegraphics[width=0.9\figwidth]{\figpath/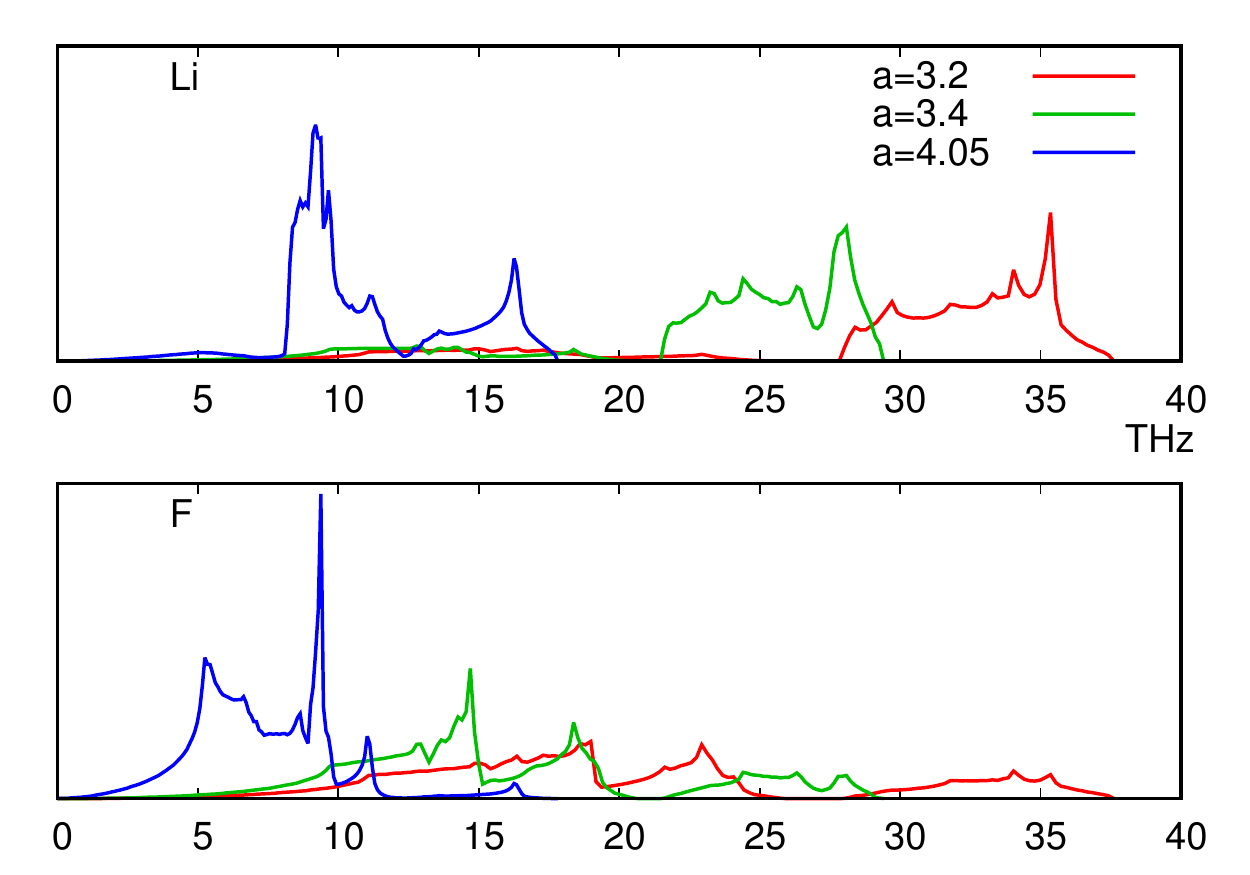}}
\subfigure[B2]
{\includegraphics[width=0.9\figwidth]{\figpath/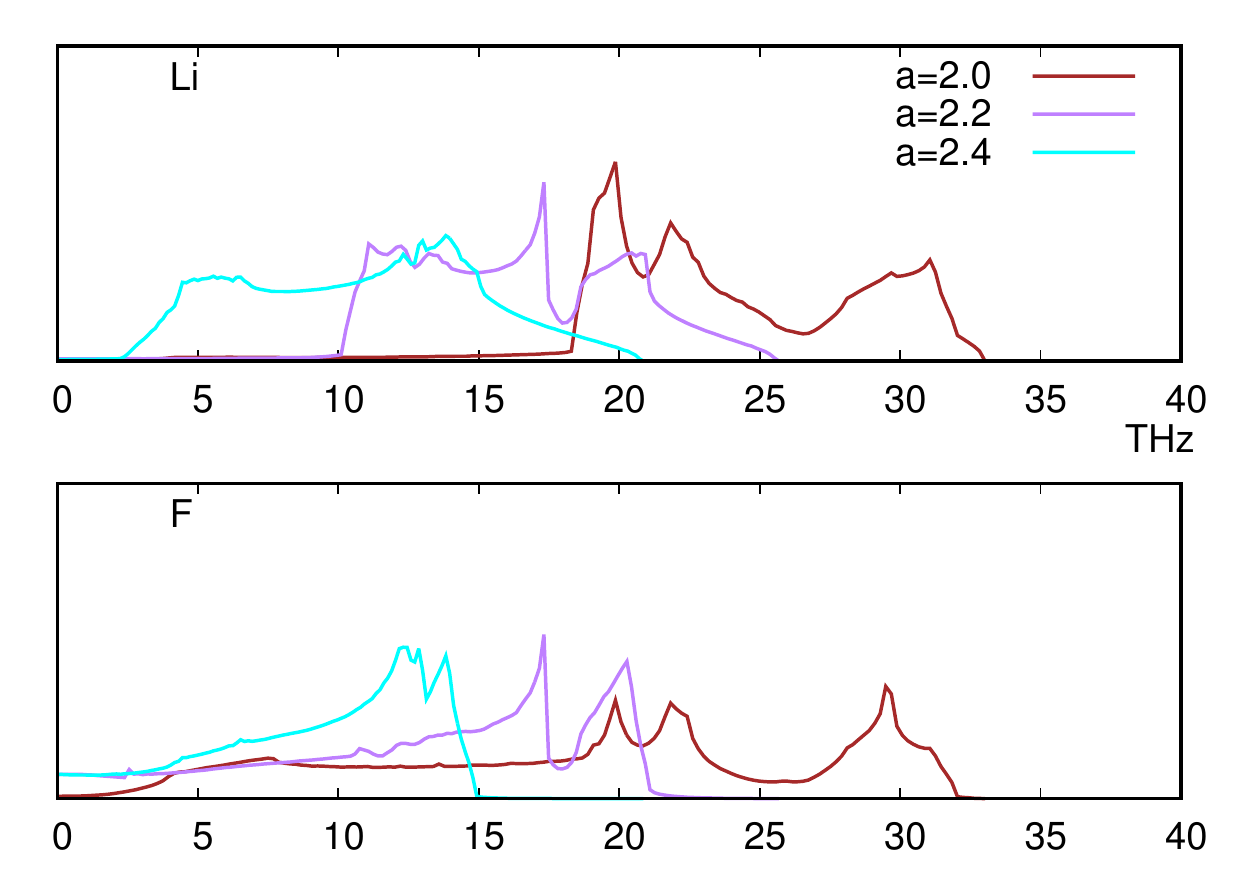}}
\caption{Dependence of the phonon density of states on deformation
\cf  \cref{dolling1968lattice} (Fig. 6).
The phDOS peak location is sensitive to strain through the dynamical matrix.
Also note that the fact that the phDOS is non-zero at zero frequency for B2:F indicates the existence of modes with negative frequencies and hence mechanical instability at $T$=0 K.
For reference, B1: lattice constants $a$=3.2, 3.4, 4.0 \AA, correspond to atomic densities $n$ = 0.24, 0.20, 0.13 atom/\AA$^3$, and B2: $a$=2.0, 2.2, 2.4 \AA, correspond to $n$ = 0.25, 0.19, 0.14 atom/\AA$^3$, where $a_\text{B1} = \sqrt[3]{4} a_\text{B2}$ gives the same density $n$. 
}
\label{fig:pdos}
\end{figure}

\begin{figure}[h!]
\centering
{\includegraphics[width=\figwidth]{\figpath/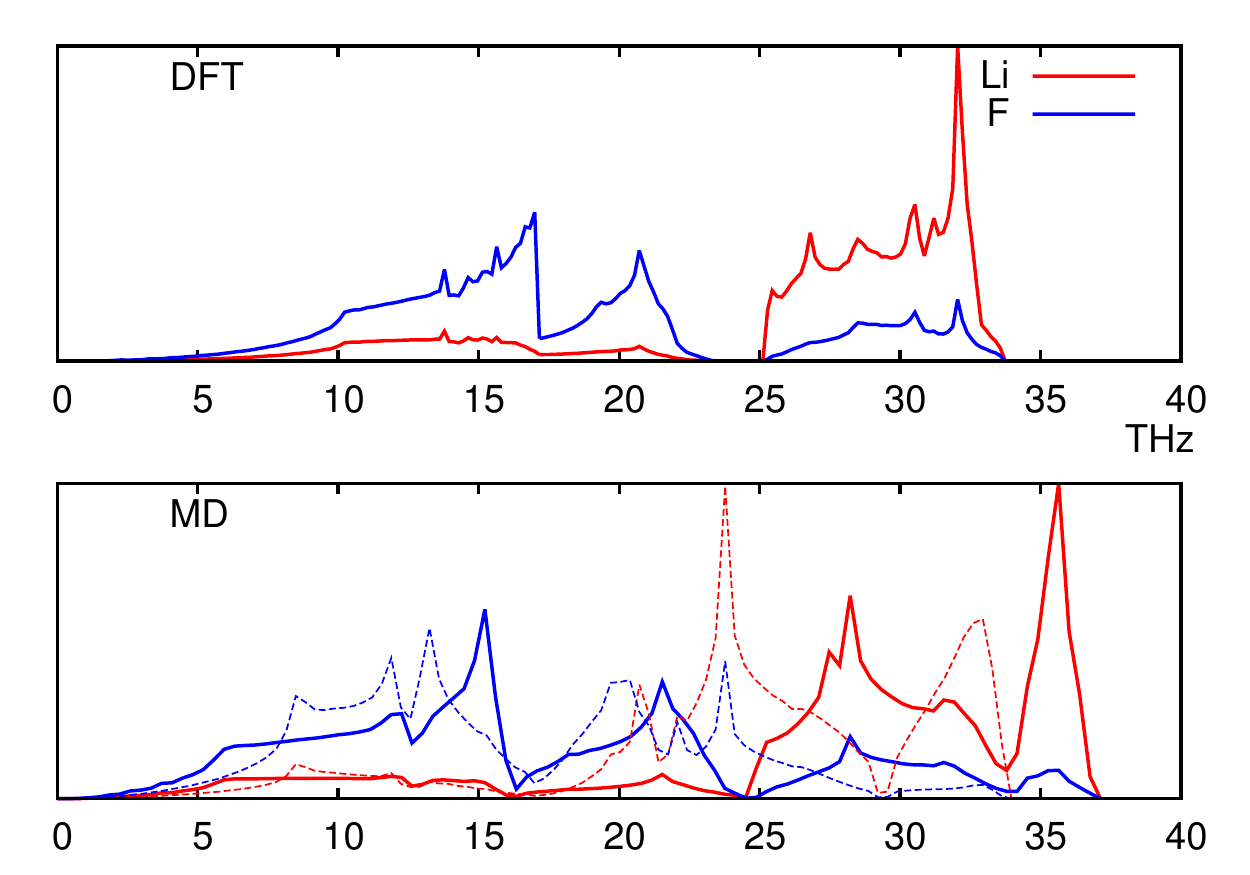}}
\caption{Comparison of MD and DFT phonon partial density of states for the B1 structure with lattice constant $a$=3.285 \AA\ which corresponds to 200 GPa and 2000 K (based on the Belonoshko potential).
In the lower phDOS the solid lines are the results for the modified TS/BMH potential where the well-depth parameter for the Li-F interaction has been increased by 30\%  (refer to \tref{tab:parameters}) relative to the original Belonoshko parameterization (dashed lines) and compare well to the upper \abinitio phDOS.
}
\label{fig:pdos_comparison}
\end{figure}

\fref{fig:pdos} shows that the phDOS (calculated via the DFT code VASP \cite{kresse1999ultrasoft,vasp} and the phonopy package \cite{togo2015first,phonopy}) is quite sensitive to compression and hence pressure.
The presence of negative frequencies in the phDOS of compressed B2 structures also indicates that the B2 phase is unstable for lattice constant $a > $ 2.1 \ \AA\ (and number density $n = N/V < $ 0.215 \AA$^3$).
We compared the phDOS for simulations with 2$\times$2$\times$2, 3$\times$3$\times$3, and 4$\times$4$\times$4 unit cells with the correction based on Born effective charges \cite{phonopy} and found results essentially indistinguishable and hence we employed systems with 2$\times$2$\times$2 for the remainder of the calculations.
For the comparison of the phDOS derived from the Belonoshko potential and that from DFT, we picked the compressed B1 configuration with $a$= 3.285 \AA\ (corresponding to 2000 K, 200 GPa lattice constant based on the Belonoshko parameterization) as representative of our pressure-temperature region of interest.
Given the poor match shown in \fref{fig:pdos_comparison}, we re-tuned the potential to achieve a better correspondence, particularly of the peaks in the phDOS, which is also shown in \fref{fig:pdos_comparison}.
Note that only changing Li-F well depth resulted in stable modifications of the crystal that maintained a reasonably representative lattice constant:  B1: 4.051 \AA\ (original) \vs 4.206 \AA\ (modified), and B2: 2.514 \AA\ (original) \vs 2.588 \AA\ (modified), at zero temperature.
The resulting and original parameters are given in \tref{tab:parameters}.

\begin{table}
\centering
\begin{tabular}{l | r r | r r}
      &  A [eV] & B [{\AA}$^{-1}$] 
      &  A' [eV] & B' [{\AA}$^{-1}$] \\ 
\hline
\hline
Li-Li &  98.933 & 3.3445 & " & " \\
Li-F  & 401.319 & 3.6900 & 521.714  & " \\
Li-Li & 420.463 & 3.3445 & " & " \\
\hline
\end{tabular}
\caption{Potential parameters of the Tosi-Fumi/Born-Mayer-Huggins style potential \eref{eq:Born}: A,B from \cref{Belonoshko2000Born} and modified A', B' based on matching an \abinitio phDOS, see \fref{fig:pdos_comparison}.
Note in \cref{Belonoshko2000Born} the short-range parameters C and D are set to zero for high pressure stability. 
Also unit charges are assigned to Li ($z_\text{Li}=$+1) and F ($z_\text{F}=$ -1).
} 
\label{tab:parameters}
\end{table}

Regarding the possibility that LiF can have B1, B2, and liquid phases over the pressure--temperature range of interest and these structural changes can influence the thermal conductivity, we adopted the approach to: (a) use DFT to predict the appropriate phase for given stress and temperatures conditions, and (b) use this phase to initialize the MD simulation which generally stays in this phase even if it is only meta-stable with respect to the empirical potential.
We did observe some deviations from this assumption, including defect formation and melting, which are noted in the Results section.
To this end, the QH model \eqref{eq:QH} derived from the same dynamical matrix used to generate phDOS was employed to estimate the relative free energy $\Delta G = G_\text{B2} - G_\text{B1}$ and thus the thermodynamic stability of the B2 phase relative to B1.
This model is built upon direct \abinitio estimates of the zero temperature enthalpy and limited in its range of validity by the mechanical stability of the phonon population at each particular deformation state.

\begin{figure}[h!]
\centering
{\includegraphics[width=\figwidth]{\figpath/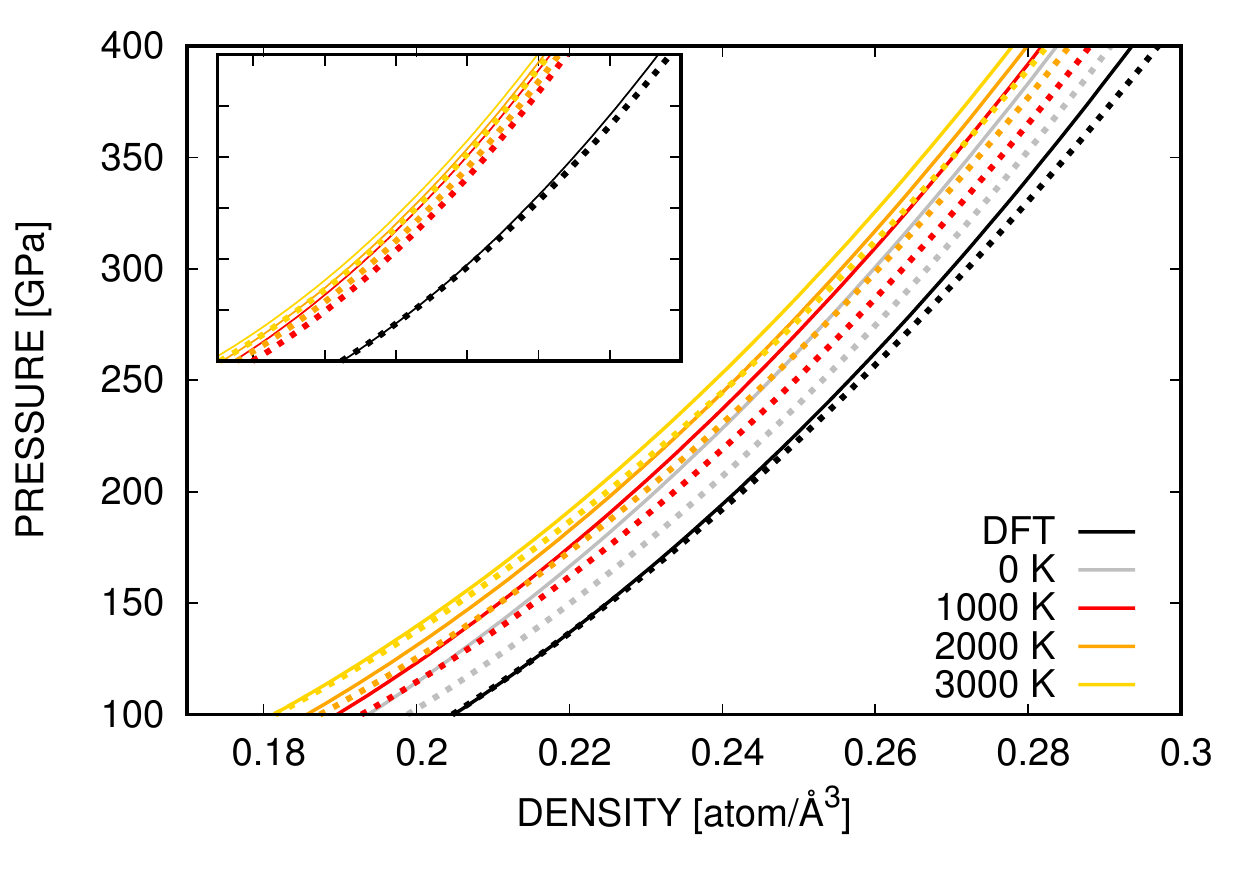}}
\caption{
Pressure as a function of (volumetric) compression and temperature for B1 (solid lines) and B2 (dashed lines) phases.
The 0K (grey) contours are  extrapolated from the higher temperature data using the linear thermal expansion model, \eref{eq:thermal_expansion}.
This extrapolation is parallel but not coincident with the DFT data (black) which implies that the MD and DFT models of LiF have similar elastic elastic properties but different zero-temperature equilibrium lattice constants.
The inset shows corresponding data for the modified MD potential.
For reference, the Rose-Vinet fits to the DFT data are:
B1: $K$= 83.93 GPa, $K'$= 4.594 GPa, $a$= 3.905 \, \AA, and
B2: $K$= 78.42 GPa, $K'$= 4.818 GPa, $a$= 2.464 \, \AA, where $K$ is the zero-temperature bulk modulus, and $a$ is the corresponding lattice constant.
For reference, the measured, ambient lattice constant is 4.03 \AA \ \cite{carrison1971compressibility}; 
}
\label{fig:pressure_density}
\end{figure}

To construct the B1--B2  phase diagram, first we constructed an equation of state.
We interpolated the function $\Pb_0$ for B1 and B2 directly from DFT data and estimated the thermal expansion coefficient  $\tcoef$ from MD data (as opposed to from the QH model) due to its full representation of anharmonic effects and its good correlation with measured values.
\fref{fig:pressure_density} shows the relevant pressure versus density curves for a range of temperatures. 
Clearly, the finite-temperature MD pressure curves are offset from the zero-temperature DFT data so that positive thermal expansion coefficients are obtained and the bulk modulus for the MD model are effectively the same as for the DFT; however, the zero-temperature equilibrium lattice constants do differ slightly.
Our estimates of the expansion coefficient $\tcoef$ employed in \eref{eq:thermal_expansion} are: B1: 0.008056 GPa/K, B2: 0.01164 GPa/K, for the original Belonoshko parameterization, and B1: 0.007411 GPa/K, B2: 0.01082 GPa/K, for the modified parameterization.
For reference, the measured coefficient of thermal expansion ($\tcoef$ divided by the bulk modulus) is 37 $\times$ 10$^{-6}$ /K \cite{combes1951mechanical} at ambient conditions which corresponds to our estimate, 30 $\times$ 10$^{-6}$ /K,  for the unmodified B1 potential.
Also apparent is the fact that the B1 and B2 phase have similar mechanical responses with B2 being slightly, but distinctly softer than B1 at the same (number) density $n$.

\begin{figure}[h!]
\centering
{\includegraphics[width=\figwidth]{\figpath/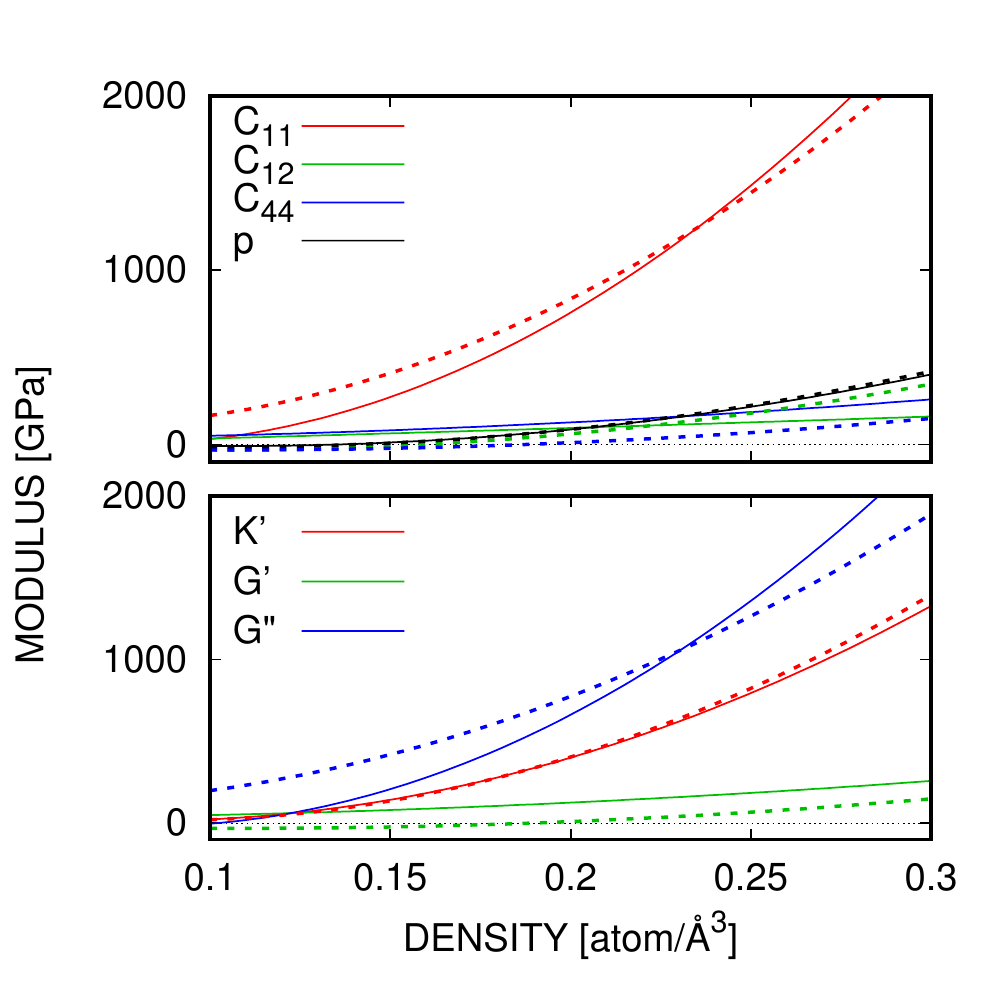}}
\caption{
Volumetric compression: elastic $C_{11}$, $C_{12}$, $C_{44}$, 
and stability moduli 
$K' = \frac{1}{3} \left( C_{11} + 2 C_{12} \right) - p$, 
$G' = C_{44}$, 
$G''= C_{11}- C_{12}$
where $C_{ij}$ are Voigt moduli about the deformed configuration.
B1 (solid) is elastically stable over the given range and B2 (dashed) is stable for $n > 0.2 $ atoms/\AA$^3$
}
\label{fig:stability}
\end{figure}

Next, we ascertained the mechanical stability of the B1 and B2 phases through the \abinitio estimates of the elastic moduli and phDOS.
\fref{fig:stability} shows the pressure and elastic moduli calculated from the DFT data, and the derived stability moduli (see \aref{app:stability} for details).
The results: (a) B1 is stable over the high pressure range we consider, and (b) B2 is only conditionally stable ($a <$ 2.15 \AA , $n >$ 0.2 atoms/\AA$^3$) based on linearized, long-wavelength elastic stability considerations, are comparable to the findings in \cref{Smirnov2011abinitio}(Fig. 2).%
\footnote{
Smirnov \cite{Smirnov2011abinitio} apparently omits the pressure dependence despite deriving the pressure dependence of the stability moduli in an earlier co-authored publication \cite{sinko2002ab}. 
Nevertheless, the shear stability criteria used in \cref{Smirnov2011abinitio} coincide with those derived in \aref{app:stability}.}
Examining phonon spectrum corresponding to the phDOS data in \fref{fig:pdos} gives a more detailed account of stability since each mode can be examined independently (see \aref{app:stability} for a discussion of the connection between the two stability criteria).
From the phDOS data, B2 is apparently stable for $a <$ 2.1 \AA, which corresponds approximately to pressure $p\approx$ 120-130 GPa for the temperatures we consider.
(Coincidently, Belonoshko \cite{Belonoshko2000Born} speculates that a  B1--B2 transition occurs  at approximately 130 GPa, which is in the neighborhood of $n$ = 0.2 atoms/\AA$^3$ given \fref{fig:pressure_density}.)

\begin{figure}[h!]
\centering
{\includegraphics[width=\figwidth]{\figpath/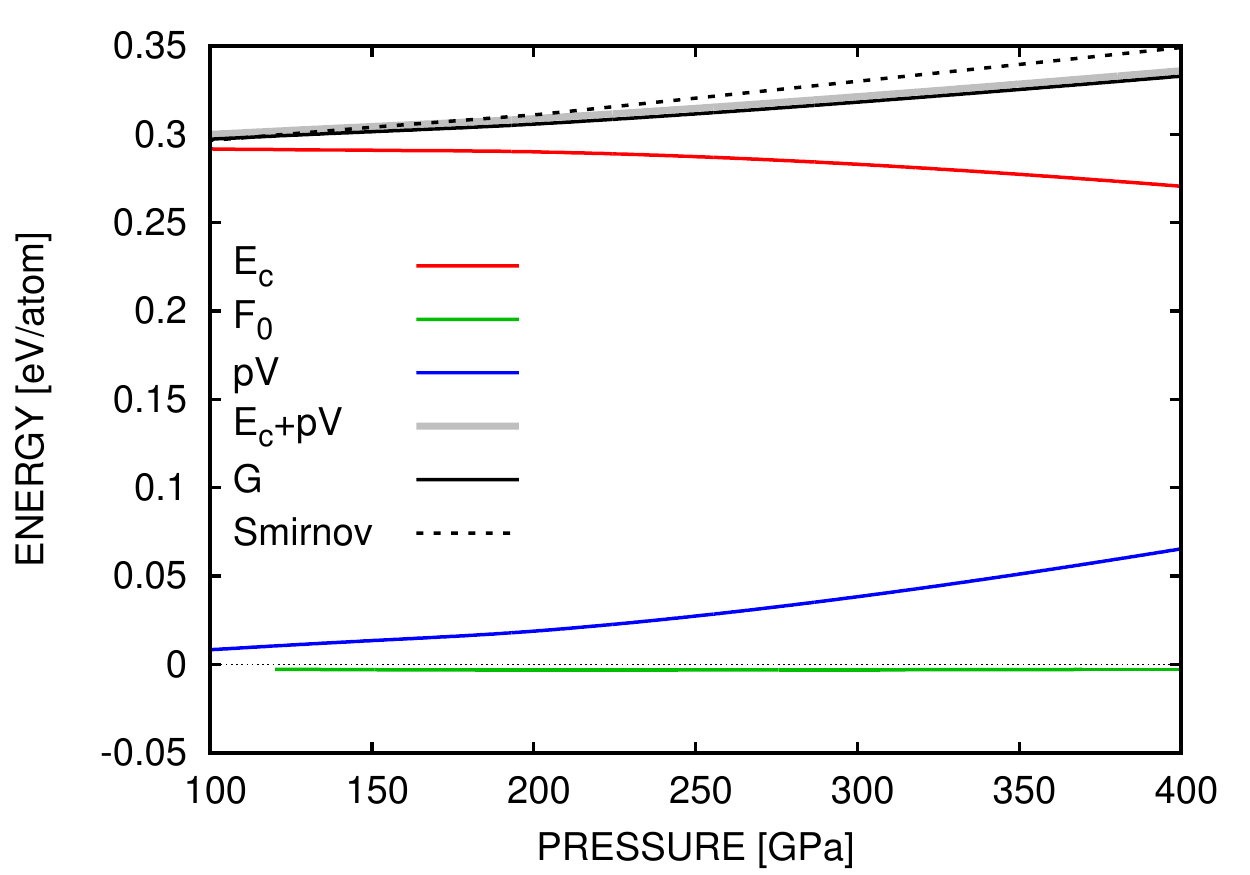}}
\caption{Energy differences (B2 relative to B1) at zero temperature calculated with DFT, \cf \cref{Smirnov2011abinitio}(Fig.1).
The results are nearly the same up to zero point energy difference $\Delta F_0$, which is nearly constant at 0.02 eV/atom over the pressure range, which is omitted in Smirnov's estimate of the Gibbs free energy difference $\Delta G$.
Note the zero point energy difference curve stops at the point of B2 instability. 
Here $E_c$ denotes cohesive energy, $F_0$ the zero-point phonon energy, $p V$ pressure-volume work and $G$ the Gibbs free energy.
}
\label{fig:enthalpy}
\end{figure}

\begin{figure}[h!]
\centering
{\includegraphics[width=\figwidth]{\figpath/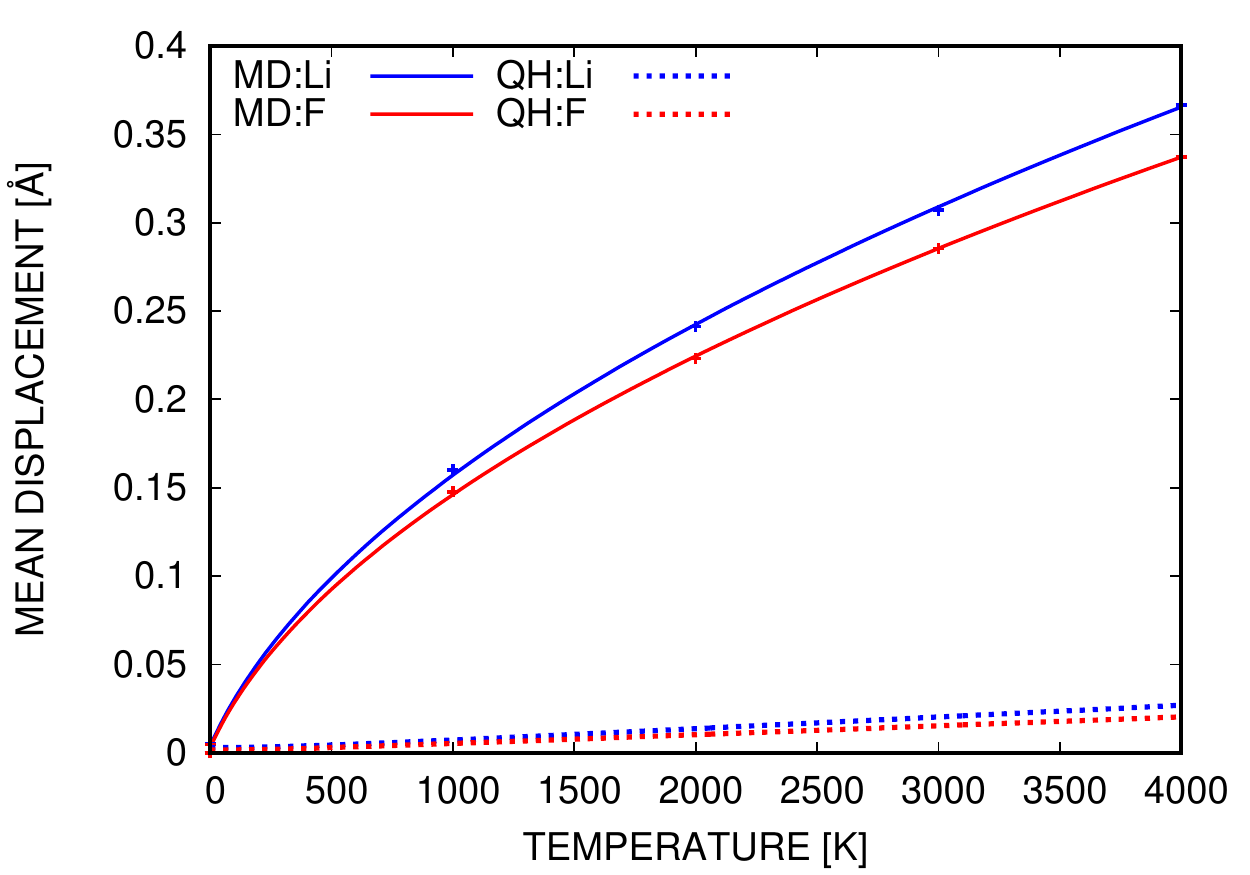}}
\caption{Mean thermal displacements for B1:$a$= 3.2 \AA. Solid lines: molecular dynamics (MD) data, dashed: \abinitio quasi-harmonic (QH) data.
}
\label{fig:thermal_displacement}
\end{figure}

Finally, we evaluated the free energy difference.
The zero-temperature energy (enthalpy) differences between B2 and B1 shown in \fref{fig:enthalpy} display trends similar those shown in \cref{Smirnov2011abinitio}(Fig.1).
Using the QH model \eref{eq:QH}, we calculate the zero point energy difference $\Delta F_0$, omitted in \cref{Smirnov2011abinitio}, to be nearly uniformly 0.02 eV/atom over the pressure range we examined, and, hence, has no significant effect on the resulting B1-B2 phase separator.
In fact, the change in zero-temperature enthalpy difference between the two phases over the given pressure range is dominated by the pressure-volume work.%
\footnote{We also calculated the enthalpy difference with the generalized gradient approximation (GGA) but it had a downward trend with pressure which we attributed to GGA tendency to over-binding.}
Unlike \cref{Smirnov2011abinitio}(Fig.6) which shows a transition to B2 at temperatures and pressures as low as 1500 K and 150 GPa, we estimate that $\Delta G > $ 0.1 eV over the given $T$ and $p$ range so B1 is always relatively thermodynamically stable.
The contours of $\Delta G$ resemble the slope of B1-B2 separator in \cref{Smirnov2011abinitio}(Fig.6), and the elastic moduli as a function of pressure and phase are similiar.
The QH model of the free-energy difference is arguably better than Debye model tuned by linear muffin tin data employed in \cref{Smirnov2011abinitio} since the QH model does not make assumptions about form of the dispersion relation; however, its validity at these temperatures is suspect.
As \fref{fig:thermal_displacement} shows, the thermal displacements predicted by the QH model are large, but still much smaller than those given by MD with similar elastic properties.
This data gives credence to the notion of thermal stabilization of apparently mechanically unstable phases. 
In this form of non-linear stability, at high temperatures atoms primarily reside outside the zero-temperature minimum state which may be mechanically unstable and in nearby regions of the energy surface with positive curvature \cite{antolin2012fast,monserrat2013anharmonic}.

\begin{figure}[h!]
\centering
{\includegraphics[width=\figwidth]{\figpath/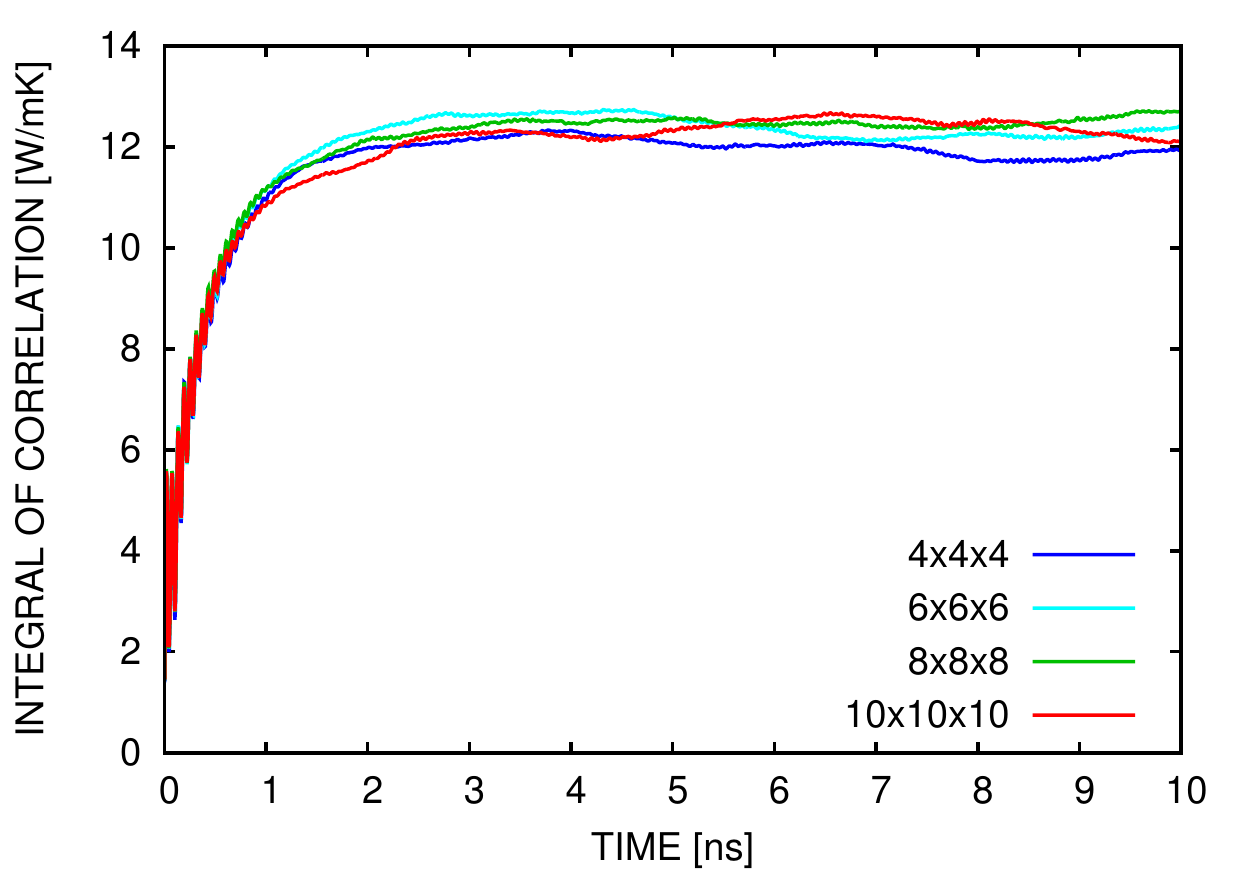}}
\caption{Sensitivity of estimated thermal conductivity to system size $V$.
Volumes: 
4$^3$  unit cells, 14.036$^3$ \AA$^3$,  512 atoms;
6$^3$  unit cells, 21.047$^3$ \AA$^3$, 1728 atoms;
8$^3$  unit cells, 28.051$^3$ \AA$^3$, 4096 atoms;,  
10$^3$ unit cells, 35.062$^3$ \AA$^3$, 8000 atoms.
}
\label{fig:size}
\end{figure}

After these validation and phase determination procedures, 
we thermalized and pressurized LiF lattices with a Nos\'e-Hoover thermo-barostat (using the classical MD code LAMMPS \cite{lammps})
in order to obtain the equilibrium flux correlations necessary to estimate thermal conductivity.
After equilibration, we used 10 replica systems with initial conditions selected from the constant temperature-pressure equilibration simulations of constant energy dynamics to compute the average flux correlation. 
After transients due to the relaxation from constant temperature dynamics subsided, samples of the correlations contributing to the average $\langle \Jb(0) \otimes \Jb(t) \rangle$ were collected every ten 0.5 fs time-steps  
As a last preliminary, given the spatial decomposition employed by PPPM, we checked for finite size effects in the flux correlations.
\fref{fig:size} shows that they are negligible with respect to the inherent noise even for periodic systems as small as 4$\times$4$\times$4 unit cells;
hence, in the following studies we used 4$\times$4$\times$4 systems and a real space/$\kb$-space decomposition cutoff $1$ nm for the PPPM electrostatic solver.

\section{Results} \label{sec:results}


Using the Green-Kubo (GK) method described in \sref{sec:theory} and the preliminaries given in \sref{sec:method}, we compute the thermal conductivity for compressed states 
in two studies for different deformations of the lattice: (a) volumetric compression over a range of pressures $p$ = 100--400 GPa and temperatures $T$= 1000--4000 K, and (b) uniaxial compression with  normal stress 1--50 GPa to simulate conditions at the initiation of a ramp experiment.
Since, at ambient pressure the measured Debye temperature for LiF is 732 K \cite[Table 12.1]{piroth2007fundamentals} and the melt is temperature 1121 K \cite{douglas1954lithium}, 
our conditions are well within the classical regime and some of the states may melt.

In preliminary studies, we found the difference in the estimated conductivity with the  modified \vs the original Belonoshko parameters was at most 10\% over the range of interest and typically only 3\%.
Since these differences were comparable to our error estimated from 10 replicas, we report conductivities derived from the original parameters.
We attribute these small differences between parameterizations with distinct phDOS, and hence dispersion characteristics, to the observation that low frequency/long wavelength phonons carry most of the heat and in that range the phDOS of the two parameterizations agree fairly well.
In fact,  G. Chen and co-authors \crefs{dames2006thermal,esfarjani2011heat} showed that 90\% of the heat in Si at ambient conditions is carried by phonons with frequencies less than about 2 Thz (estimated from the reported 2-5 nm wavelength and the given elastic moduli).

\subsection{Volumetric compression}

\begin{figure}[h!]
\centering
{\includegraphics[width=\figwidth]{\figpath/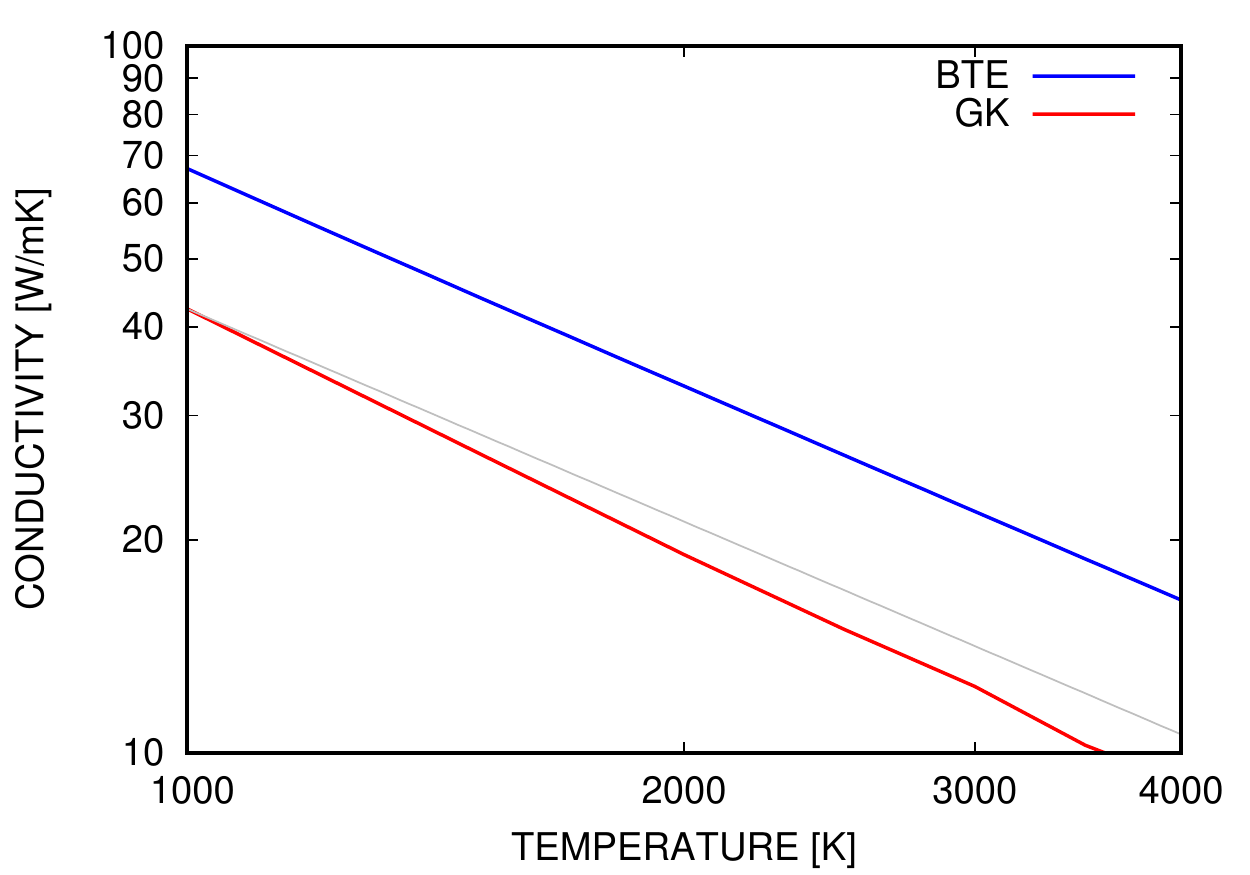}}
\caption{Comparison of \abinitio BTE and MD GK on a log-log scale for B1:$a$= 3.2 \AA.
The grey trendline is $k \sim T^{-1}$.
}
\label{fig:abinitio_comparison}
\end{figure}

First, we compared the conductivity estimated with MD GK and a method directly employing \abinitio data for B1 LiF at constant volume ($a$= 3.2 \AA) over the temperature range $T$=1000--4000 K. 
Specifically, in the second method the Boltzmann transport equation (BTE) was parameterized with \abinitio second and third order force constants derived from 2$\times$2$\times$2, and 4$\times$4$\times$4 unit cell systems respectively and solved in the single mode relaxation time approximation.
See \crefs{togo2015distributions,phono3py} for details, and the similar approach in \cref{li2014shengBTE}.
Given the differences in the methods, the results shown in \fref{fig:abinitio_comparison} are comparable.
The MD estimates are uniformly lower than those of the BTE, which is consistent with fact that the MD has a complete, albeit less exact, Hamiltonian with no truncation of the phonon scattering interactions and the temperature is high enough for higher order and non-linear mechanisms beyond those captured by a single relaxation time to be significant.
Also noteworthy, the thermal conductivity derived from the BTE model displays perfect $T^{-1}$ scaling, whereas the MD estimates show slightly stronger decay with temperature.

\begin{figure}[h!]
\centering
{\includegraphics[width=1.5\figwidth]{\figpath/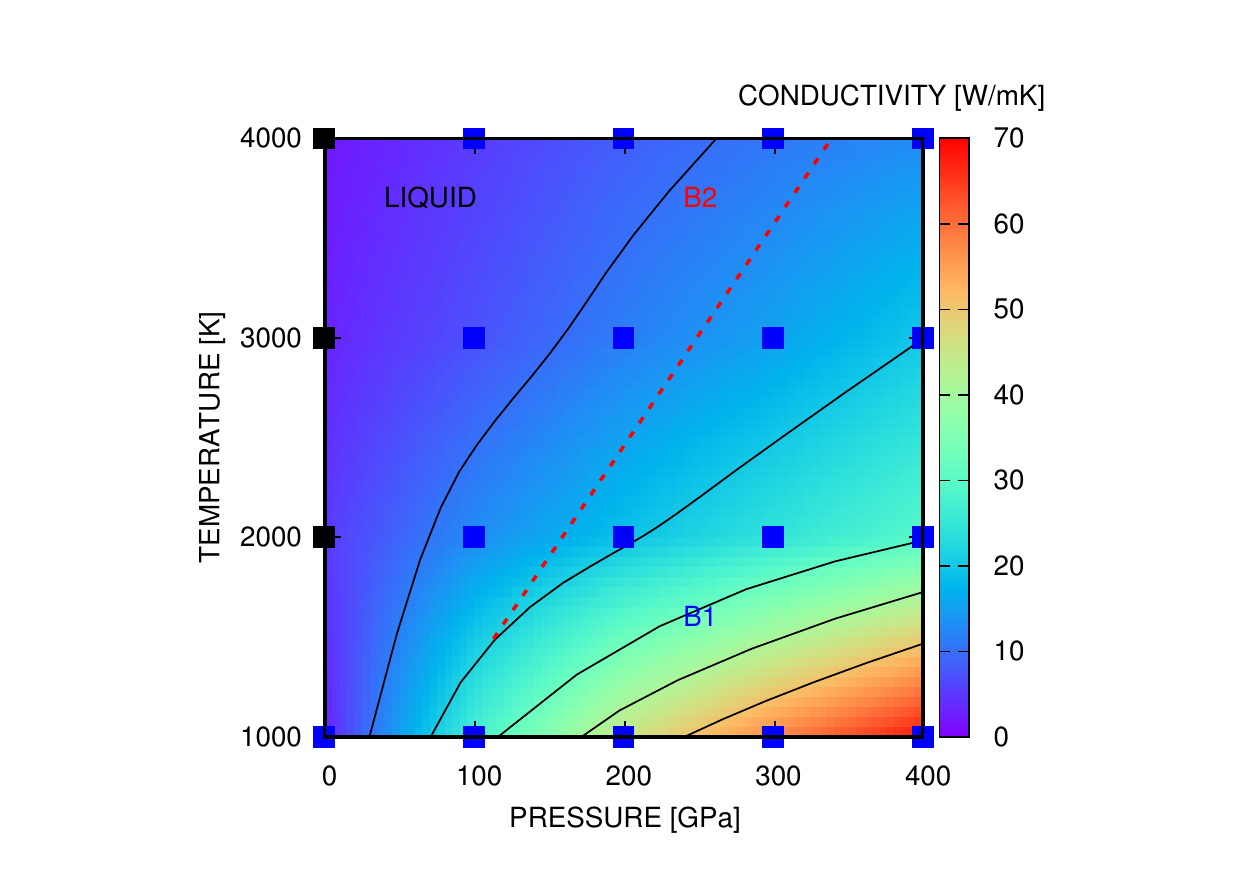}}
\caption{Volumetric deformation: thermal conductivity $\kappa$ [W/K-m] as a function of pressure P [GPa] and temperature T [K], and crystal structure.
The phase of each sample is indicated by the color of the square data points: B1:blue, B2:red, melt:black. 
The red dashed line corresponding to the B1-B2 transition calculated by Smirnov \cite{Smirnov2011abinitio} is shown for reference.
}
\label{fig:vol_cond}
\end{figure}

\fref{fig:vol_cond} shows the thermal conductivity estimated with MD GK for pressures in the range 1--400 GPa and temperatures 1000--4000 K, and \tref{tab:kappa} gives the corresponding data for both the B1 and B2 phases.
In \fref{fig:vol_cond} the phase of the samples used to create the contour plot are marked and a few of the high temperature, relatively low pressure systems melted.
As can be seen in \tref{tab:kappa}, the estimated thermal conductivity for the B1 and B2 phases have comparable values and same trends.
This is plausible given that the elastic properties of the two phases are similar and a simple kinetic model of thermal transport indicates that the resulting comparable sound speeds should lead to similar conductivities.
The same basic kinetic interpretation is consistent with the observations that the thermal conductivity increases with increased pressure due to higher wave speeds and with lower temperature due to relatively less scattering and longer phonon mean free path. 
These trends are  monotonic and have decreasing effect on the thermal conductivity.

Although the Belonoshko potential was tuned to high pressure conditions, we also calculated the thermal conductivity nearer ambient conditions. 
The values we obtain, \eg 2.8$\pm$0.2 W/mK at 1 GPa, 1000 K and 1.8$\pm$0.2 W/mK at 1 atm, 1200 K (melt, refer to \tref{tab:kappa}) are roughly comparable to the value 1.5 W/mK at 1 atm, 1150 K (melt) given by Ishi \etal \cite{Ishii2014conductivity} using a different empirical potential and the experimental measurements:
15.7 W/mK at 0.1 GPa,  
16.3 W/mK at 1.0 GPa, 300 K \cite{andersson1987thermal}, 
and
 4.0 W/mK at 1 atm, 314 K \cite{combes1951mechanical} of solid LiF.

\begin{table}
\centering
\subtable[] {
\begin{tabular}{r r | r r }
$p$      &    T     & $\kappa_\text{B1}$   & $\kappa_\text{B2}$ \\
\hline
\hline
 1     &   1000  &  2.83 &  -  \\
       &   2000  &  {\it 5.13} &  -  \\
       &   3000  &  {\it 3.15} &  -  \\
       &   4000  &  {\it 2.01} &  -  \\
\hline
\end{tabular}
}
\subtable[] {
\begin{tabular}{r r | r r }
$p$      &    T     & $\kappa_\text{B1}$   & $\kappa_\text{B2}$ \\
\hline
\hline
 100   &   1000  & 27.29 & {15.42}$^*$  \\        
       &   2000  & 12.93 &  {8.52}$^*$  \\
       &   3000  &  7.19 &  {\it 6.04}  \\
       &   4000  &  5.20 &  {\it 5.54}  \\
\hline
\end{tabular}
}
\subtable[] {
\begin{tabular}{r r | r r }
$p$      &    T     & $\kappa_\text{B1}$   & $\kappa_\text{B2}$ \\
\hline
\hline
200    &   1000  & 45.20 &  40.45    \\
       &   2000  & 19.16 &  19.67    \\
       &   3000  & 11.60 &  12.34    \\ 
       &   4000  &  8.47 &  {\it 6.45}    \\
\hline
\end{tabular}
}
\subtable[] {
\begin{tabular}{r r | r r }
$p$      &    T     & $\kappa_\text{B1}$   & $\kappa_\text{B2}$ \\
\hline
\hline
300    &   1000  & 57.23 &      61.79   \\ 
       &   2000  & 24.38 &      28.36   \\ 
       &   3000  & 15.37 &      18.04   \\ 
       &   4000  & 10.96 &      {\it 7.82}   \\ 
\hline
\end{tabular}
}
\subtable[] {
\begin{tabular}{r r | r r }
$p$      &    T     & $\kappa_\text{B1}$   & $\kappa_\text{B2}$ \\
\hline
\hline
400    &   1000  & 68.18 &      74.88   \\ 
       &   2000  & 29.43 &      34.20   \\ 
       &   3000  & 27.28 &      22.69   \\ 
       &   4000  & 13.06 &  {\it 8.71}  \\ 
\hline
\end{tabular}
}
\caption{Volumetric compression: thermal conductivity $\kappa$ [W/K-m] as a function of pressure P [GPa] and temperature T [K], and crystal structure (values in {\it italics} are for melted crystals).
Errors in estimated $\kappa$ are  $<$ 5 \% based on predictions from 10 replicas.
$^*$ Note for B2 at 100 GPa, 1000-2000 K approximately half of the replicas transform to twinned B2 structures in initialization.
}
\label{tab:kappa}
\end{table}

\subsection{Uniaxial compression}

\begin{figure}[h!]
\centering
{\includegraphics[width=0.9\figwidth]{\figpath/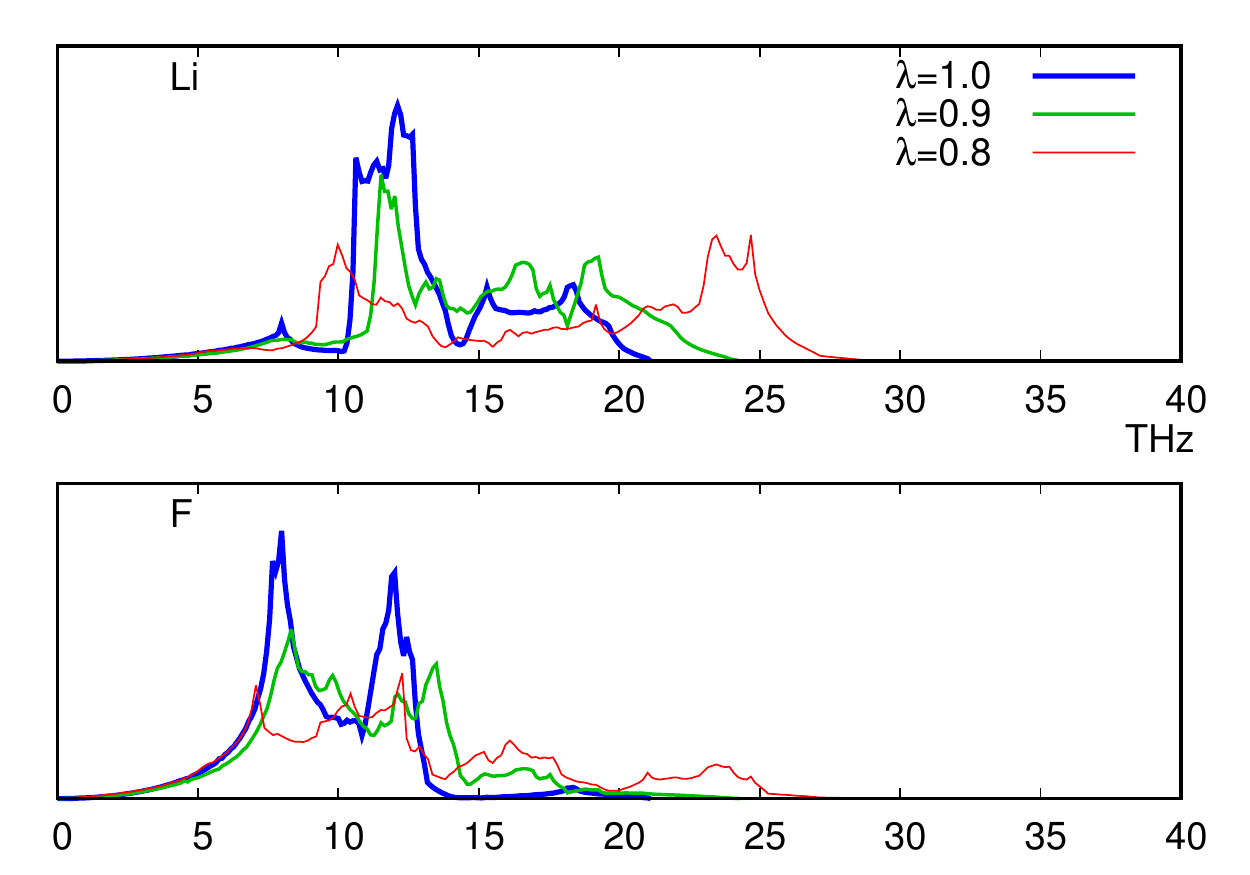}}
\caption{Uniaxial compression: phonon partial density of states dependence on deformation, where $\lambda$ is the compression in the 110 direction.
}
\label{fig:uniaxial_pdos}
\end{figure}

For this study, we compressed one direction of a B1 LiF crystal while fixing the lateral dimension to a 4.02 \AA\ lattice spacing to mimic initial phases of ramp compression with inertial confinement and examine the resulting differences in the thermal conductivity resulting from unequal principal strains.
The compression direction was chosen to be  $110$, since this direction lacks surface polarization.
The lateral directions were $1\bar{1}0$ (equivalent to $110$) and $001$, respectively.
The compressions $\lambda \in \{ 0.8, 0.9, 1.0 \}$ examined corresponded to normal stresses 1--50 GPa in the compressed dimension (note that  75 GPa MD crystal was unstable and a dislocation formed) and temperature range  1000--3000 K.
It was not possible to predict which phases were thermodynamically stable over this deformation--temperature range since we predicted that B2 has unstable phonons over the range we examined.
The phDOS for B1, \fref{fig:uniaxial_pdos}, shows that the compressed direction becomes stiffer (the sound speed is roughly inversely proportional to slope of phDOS) and higher frequency content is added to the phonon spectrum.

\begin{figure}[h!]
\centering
{\includegraphics[width=1.0\figwidth]{\figpath/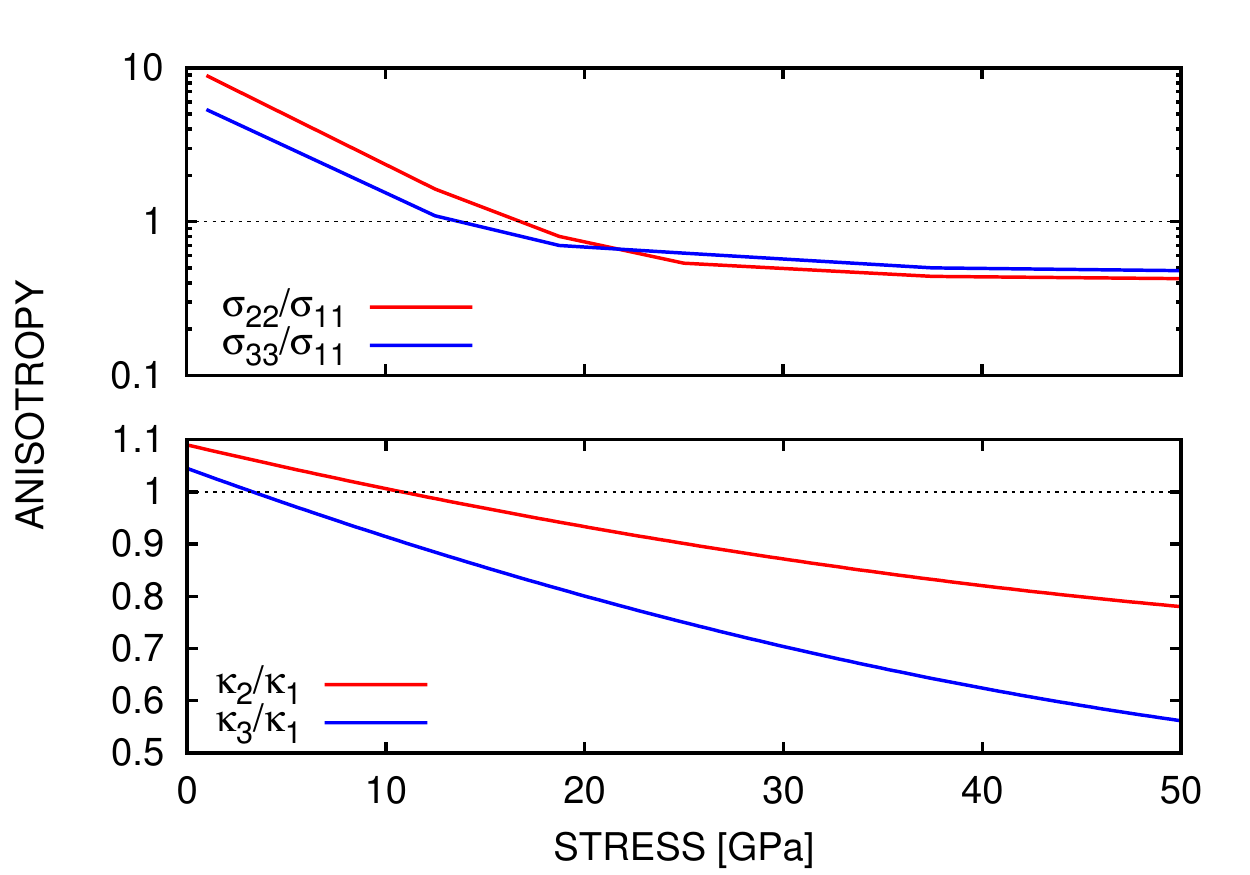}}
\caption{Uniaxial compression: stress ratios (top) and conductivity anisotropy (bottom) at $T$=1000 K and all for the B1 phase.
The coordinate axes aligned with 110, 1$\bar{1}$0 and 001 crystal directions, and the 110 direction is compressed and the others are unstrained relative to the zero temperature lattice.
}
\label{fig:uniaxial_anisotropy}
\end{figure}

\begin{figure}[h!]
\centering
{\includegraphics[width=1.5\figwidth]{\figpath/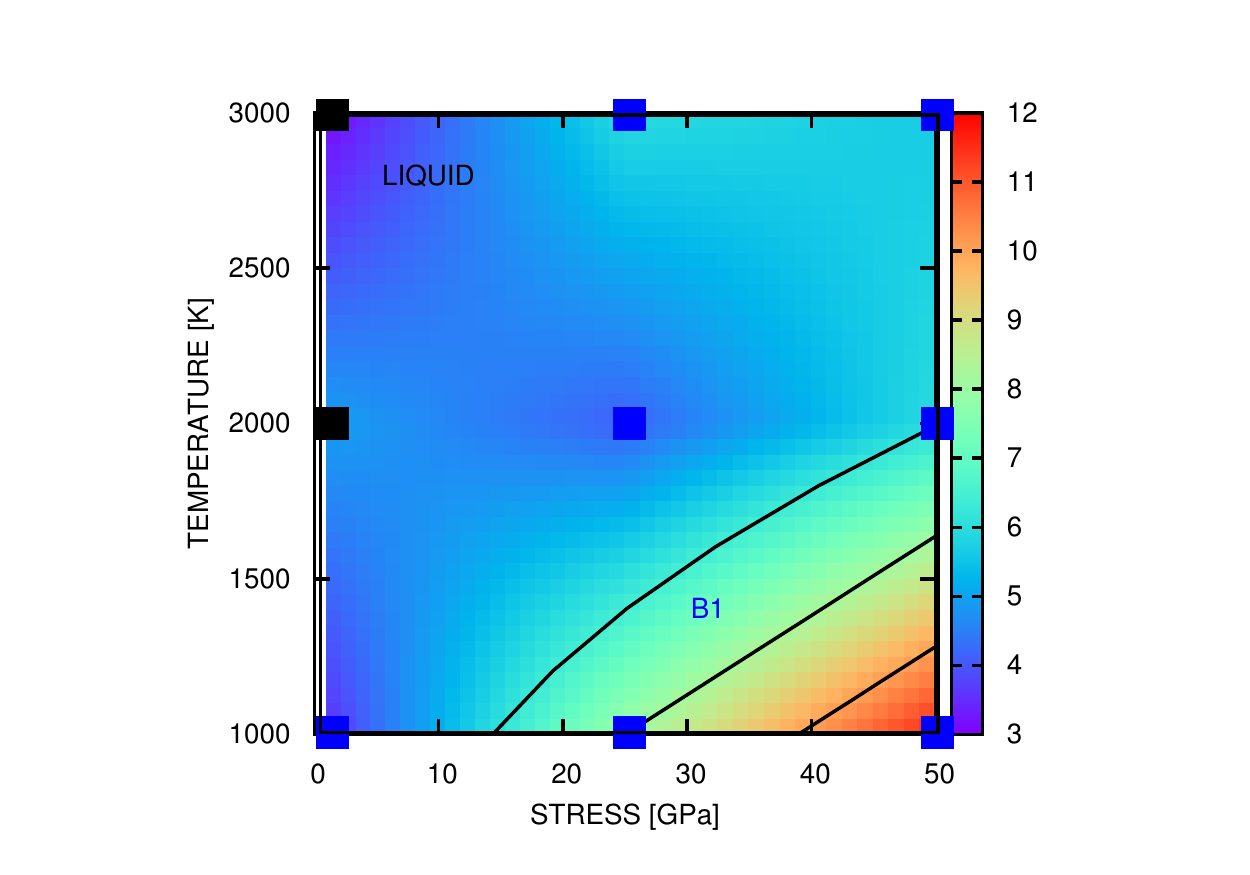}}
\caption{Uniaxial compression: thermal conductivity in the compressed 110 direction.
}
\label{fig:uniaxial_conductivity}
\end{figure}

\fref{fig:uniaxial_anisotropy} shows stress response to these conditions and corresponding anisotropy of the thermal conductivity of the B1 structure.
Note that the initial lattice constant is not equilibrium at the given temperatures which immediately induces the anisotropy shown.
Also  the lateral stresses become nearly equal but distinct from the normal stress in the compressed direction as the structure loses perfect crystallinity.
\fref{fig:uniaxial_conductivity} shows that the normal conductivity follows similar trends with temperature and pressure as in the volumetric compression case, namely in this state the highest conductivities are at the highest pressures and lowest temperatures.
The data for this study is tabulated in \tref{tab:uniaxial_kappa}.


\begin{table}
\centering
\subtable[] {
\begin{tabular}{r r r r | r r r }
$\sigma_{11}$ & $\sigma_{22}$ & $\sigma_{33}$ &    T     & $\kappa_{11}$ & $\kappa_{22}$ & $\kappa_{33}$     \\
\hline
\hline
 1 & 8 &   5 &   1000  &  3.50 &  3.78 &  3.61  \\
   & 1 &   1 &   2000  &  {\it 4.94} &  {\it 5.16} &  {\it 5.27}  \\
   & 1 &   1 &   3000  &  {\it 3.04} &  {\it 3.20} &  {\it 3.34}  \\
\hline
\end{tabular}
}
\subtable[] {
\begin{tabular}{r r r r | r r r}
$\sigma_{11}$ & $\sigma_{22}$ & $\sigma_{33}$ &    T     & $\kappa_{11}$ & $\kappa_{22}$ & $\kappa_{33}$     \\
\hline
\hline
 25 & 14 & 15 &   1000  &  7.99 &  7.20 &  6.00  \\
    & 19 & 20 &   2000  &  4.11 &  4.11 &  3.48  \\
    & 25 & 25 &   3000  &  5.98 &  5.58 &  5.84  \\
\hline
\end{tabular}
}
\subtable[] {
\begin{tabular}{r r r r | r r r }
$\sigma_{11}$ & $\sigma_{22}$ & $\sigma_{33}$ &    T     & $\kappa_{11}$ & $\kappa_{22}$ & $\kappa_{33}$  \\
\hline
\hline
 50 & 21 & 24 &   1000  & 11.59 &  9.04 &  6.51 \\
    & 50 & 28 &   2000  &  5.96 &  4.67 &  3.93 \\
    & 50 & 50 &   3000  &  5.64 &  5.51 &  5.60 \\
\hline
\end{tabular}
}
\caption{Uniaxial compression: thermal conductivity $\kappa$ [W/K-m] as a function of normal stresses $\sigma_{11}$, $\sigma_{22}$, $\sigma_{33}$ [GPa] and temperature T [K], for the B1 crystal structure (values in {\it italics} are for melted crystals).
Errors in estimated $\kappa$ are  $<$ 5 \% based on predictions from 10 replicas.
Lateral dimension set at 4.02 \AA\ lattice.
}
\label{tab:uniaxial_kappa}
\end{table}

\section{Discussion} \label{sec:discussion}

In summary, we found that the thermal conductivity of LiF at high temperatures and pressures is only marginally dependent on phase and ranged from about 5 W/mK to 70 W/mK over the range 1000--4000 K and 100--400 GPa.
For our purposes, the fact that the two expected phases (B1 and B2) have similar conductivity  offsets the difficulties in determining their mechanical and thermodynamic stability.
Our estimates are corroborated by the limited experimental data available as well as direct \abinitio estimates of thermal conductivity.
We also found that the uniaxial deformation expected to result from inertia confinement of the targeted ramp compression experiments may lead to significant anisotropy in the thermal conductivity.
More rigorous treatment of the relative phase stability exist in the literature than the method we selected, notably \cref{monserrat2013anharmonic} which focussed on the influence of the anaharmonic phonon energy and \cref{desjarlais2013first} which adapts the phase coexistence technique of \cref{lin2003two}  to finite temperature DFT calculations, which may shed light on phase transitions from B1 at high temperatures and pressures.
Since our findings indicate that these transitions are unlikely over our temperature and pressure range of interest, whereas the formation of defects appear at relatively low uniaxial compression we intend to pursue investigation of the influence of defects on the thermal conductivity of solid LiF next.

\section*{Acknowledgements}
We thank Luke Shulenberger, Catalin Spataru and Thomas Mattsson for helpful guidance and appreciate the use of LAMMPS \cite{lammps}, VASP \cite{vasp}, and phonopy \cite{phonopy}.
This work was supported by the NNSA Advanced Simulation and Computing - Physics and Engineering Models program at Sandia National Laboratories.
Sandia is a multiprogram laboratory managed and operated by Sandia Corporation, a wholly owned subsidiary of Lockheed Martin Corporation, for the U.S. Department of Energy's National Nuclear Security Administration under contract No. DE-AC04-94AL85000.

\clearpage

\appendix

\def\theequation{A.\arabic{equation}}
\setcounter{equation}{0}
\section{Virial stress with Coulomb interactions} \label{app:virial}

Although we employ the PPPM method in \sref{sec:results} the gist of how the virial stress and, hence, the heat flux is obtained is easier to explain in the context of the  Ewald sum \cite[Eq. 7]{karasawa1989acceleration}:
\begin{equation}
\begin{split}
\Phi &=
\frac{1}{2}\sum_{\alpha\neq\beta} \varphi(r_{\alpha\beta}) + \frac{q_\alpha q_\beta}{\epsilon r_{\alpha\beta}} \left( \erfc\left(\frac{r_{\alpha\beta}}{\ell}\right) + \erf\left(\frac{r_{\alpha\beta}}{\ell}\right) \right)  \\
&= \underbrace{ \frac{1}{2} \sum_{\alpha\neq\beta} \varphi(r_{\alpha\beta}) + \frac{q_\alpha q_\beta}{\epsilon r_{\alpha\beta}} \erfc\left(\frac{r_{\alpha\beta}}{\ell}\right)    }_{\text{real}: \ \bar{\Phi}(\{\xb_\alpha\})}
 + \underbrace{ \frac{2\pi}{\epsilon V} \sum_{\kb\neq\mathbf{0}} \frac{1}{\|\kb\|^2} \exp\left( -\frac{1}{4} \|\kb\|^2 \ell^2 \right) \Re \sum_{\alpha\neq\beta} q_\alpha q_\beta  \exp\left( \imath \kb\cdot\xb_{\alpha\beta} \right)   }_{\text{reciprocal}: \ {\tilde{\Phi}_\kb(\{\xb_\alpha\})}}
\end{split}
\end{equation}
where the error function, $\erf(r/\ell)$, and its complement, $\erfc(r/\ell) = 1 -\erf(r/\ell)$, play the role of a blending/cutoff function with parameter $\ell \sim \sqrt[3]{V} $, 
and $\xb_{\alpha\beta} = \xb_{\alpha} - \xb_{\beta}$ is a relative position vector.
Note we have used the Fourier transforms $\Fc_{\xb\to\kb} \left[ \sum_\alpha q_\alpha \delta(\xb-\xb_\alpha) \right] = \sum_\alpha \exp \imath \kb \cdot \xb_\alpha$
and $\Fc_{\xb\to\kb} \left[ \frac{1}{r} \erf(r/\ell) \right] = \frac{1}{\|\kb\|^2} \exp\left( -\frac{1}{4} \|\kb\|^2 \ell^2 \right) $.
It follows, after dropping the species subscripts $a,b$ for clarity, that the per-atom energy for pair potentials is
\begin{equation}
\begin{split}
\varepsilon_\alpha &= \frac{1}{2} m_\alpha \vb_\alpha \cdot \vb_\alpha + \frac{1}{2} \sum_{\beta\neq\alpha} \left( \varphi (r_{\alpha\beta}) + \frac{q_\alpha q_\beta}{\epsilon r_{\alpha\beta}} \erfc\left(\frac{r_{\alpha\beta}}{\ell}\right) \right)  \\
&+ \frac{2 \pi}{\epsilon V} \sum_{\kb \neq \mathbf{0}} 
\frac{1}{\|\kb\|^2} \exp\left(-\frac{1}{4}\| \kb\|^2 \ell^2\right) \Re
\sum_{\beta\neq\alpha} q_\alpha q_\beta \exp\left(\imath \kb \cdot \xb_{\alpha\beta}\right) 
\end{split}
\end{equation}
\cf \cref{heyes1994pressure}(Eq. 8).
Thus, the expression for the per-atom virial stress $\virial_\alpha$ \cite{karasawa1989acceleration,heyes1994pressure,sirk2013characteristics} is:
\begin{equation} \label{eq:virial}
\begin{split}
\virial_\alpha
=& - \frac{1}{2V} \sum_{\beta}  \left[
- \frac{\mathrm{d}}{\mathrm{d}r} \varphi(r_{\alpha\beta}) 
+ q_\alpha q_\beta \sum_{\kb\neq \mathbf{0}} \frac{2}{\sqrt{\pi}\ell} r_{\alpha\beta} \exp\left(-\frac{r_{\alpha\beta}^2}{\ell^2}\right) + \erfc\left(\frac{r_{\alpha\beta}^2}{\ell^2}\right)  
\right] \frac{1}{r_{\alpha\beta}^3}  \xb_{\alpha\beta} \otimes \xb_{\alpha\beta}   \\
&- \frac{2\pi}{\epsilon V} \sum_{\kb\neq\mathbf{0}} \left| \sum_\alpha q_\alpha \exp \imath \kb\cdot\xb_\alpha \right|^2 \frac{1}{\| \kb \|^2}{\exp\left( -\frac{1}{4}{\| \kb \|^2 \ell^2}\right)} \left[ \Ib - \left( \frac{2}{\| \kb \|^2} + \frac{1}{2} \ell^2 \right) \kb \otimes \kb \right]
\end{split}
\end{equation}
where $\Ib$ is the identity tensor and  $\frac{\mathrm{d}}{\mathrm{d}r} \erf(r/\ell) = \frac{2}{\sqrt{\pi} \ell} \exp\left(-r^2/\ell^2\right)$, \cf \cref{heyes1994pressure}(Eq. 22).

\def\theequation{B.\arabic{equation}}
\setcounter{equation}{0}
\section{Elastic moduli and stability} \label{app:stability}

Many versions of the elastic moduli tensor exist at finite deformations like those investigated in this study and the elastic stability of crystal lattices and elastic materials has been well studied, see, \eg,
\crefs{hadamard1922lectures, born1954dynamical, knops1973theory, milstein1979theoretical, milstein1979divergences, wang1995mechanical, wallace1998thermodynamics, morris2000internal, clatterbuck2003phonon, vanvliet2003quantifying, weinan2007cauchy, miller2008nonlocal, delph2009local, delph2010prediction, mouhat2014necessary}.
To connect continuum, elastic stability to atomic, phonon stability,
we will assume the current positions $\xb_\alpha$ are given by small time-varying displacements $\ub_\alpha$ due to phonon modes superposed on large, static deformations characterized by a homogeneous deformation of the zero temperature, equilibrium lattice $\Fb \Xb_\alpha$
\begin{equation}
\xb_\alpha(t) = \Fb \Xb_\alpha + \ub_\alpha(t)
\end{equation}
Since a homogeneous deformation maintains equilibrium, $\fb_\alpha(\Fb \Xb_\alpha) = \mathbf{0}$, and hence the (linearized) Newton equation governing the phonon modes is
\begin{equation}
m_\alpha \ddot{\ub}_{\alpha} = \sum_\beta \Kbb_{\alpha\beta} \ub_\beta
\end{equation}
Likewise, in the continuum limit, such that $\xb = \Fb \Xb + \ub$, the linearized balance of momentum
\begin{equation} \label{eq:bmom0}
\rho_0 \ddot{\ub} = \grad_\Xb \cdot( \Bbb \grad_\Xb \ub )
= \sum_{AjB} \Bbb_{iAjB} u_{j,AB} \eb_i
\end{equation}
governs the long-wavelength elastic waves.
Here, $\rho_0$ is the mass density in reference configuration $\Xb$.
Since background stress $\bar{\Pb} = \Pb(\Fb)$ is homogeneous the system is also in equilibrium at the continuum level.
The elasticity tensor $\Bbb$ of the first Piola-Kirchoff stress $\Pb$ with respect to the deformation gradient has an atomic-level definition
\begin{equation}
\begin{split}
\Bbb =& \frac{\partial}{\partial\Fb} \Pb = \frac{1}{V_0} \frac{\partial^2 \Phi}{\partial\Fb \partial\Fb}  
= \frac{1}{V_0} \sum_{\alpha,\beta} \left[ \frac{\partial^2 \Phi}{\partial \xb_\alpha \partial \xb_\beta}\right]_{ij} \eb_i \otimes \Xb_\alpha \otimes \eb_j \otimes \Xb_\beta
= \frac{1}{V_0} \sum_{\alpha,\beta} \left[ \Kbb_{\alpha\beta} \right]_{ij} \eb_i \otimes \Xb_\alpha \otimes \eb_j \otimes \Xb_\beta \\
\end{split}
\end{equation}
where $\Xb_{\alpha\beta} \equiv \Xb_\alpha - \Xb_\beta$. 
Further manipulation leads to
\begin{equation}
\begin{split}
\Bbb 
=& \frac{1}{V_0} \sum_{\alpha,\beta} \bigg[ 
\frac{1}{\| \Fb \Xb_{\alpha\beta} \|^2 } \left( \frac{\partial^2\Phi}{\partial \xb_{\alpha\beta}^2}  - \frac{1}{\| \Fb \Xb_{\alpha\beta} \|} \frac{\partial \Phi}{\partial \xb_{\alpha\beta} } \right) \, \Fb \Xb_{\alpha\beta} \otimes \Xb_{\alpha\beta} \otimes \Fb \Xb_{\alpha\beta} \otimes \Xb_{\alpha\beta} \\
&+ 
\frac{1}{\| \Fb \Xb_{\alpha\beta} \| } \frac{\partial\Phi}{\partial \xb_{\alpha\beta}}  \, \sum_{i=1}^3 \eb_i \otimes  \Xb_{\alpha\beta} \otimes \eb_i \otimes \Xb_{\alpha\beta} 
\bigg]
\\
=& \frac{1}{V_0}
\sum_{A,B,C,D,i,j=1}^3 \left[ \Cbb_{ABCD} \Fb_{iA} \Fb_{jC} + \Sb_{BD} \delta_{ij} \right] 
\eb_i \otimes \Eb_B \otimes \eb_j \otimes \Eb_D
\end{split}
\end{equation}
relating $\Bbb$ to the more familiar elasticity tensor 
\begin{equation}
\Cbb = \frac{\partial}{\partial\Eb} \Sb = \frac{\partial^2 \Phi}{\partial\Eb \partial\Eb} 
= \frac{1}{4} \sum_{(\alpha\beta),(\gamma\nu)} \frac{\partial^2 \hat{\Phi}}{\partial r^2_{\alpha\beta} \partial r^2_{\gamma\nu}} \Xb_{\alpha\beta} \otimes \Xb_{\alpha\beta} \otimes \Xb_{\gamma\nu} \otimes \Xb_{\gamma\nu}
\end{equation}
of the symmetric second Piola-Kirchoff stress $\Sb$ with respect to the Lagrange strain $\Eb = \frac{1}{2} \left( \Fb^T \Fb - \Ib \right)$ \cf \cref{weiner2012statistical}(Eq. 4.6.11).
Using the chain rule $u_{j,AB} = F_{kA} F_{lB} u_{j,kl}$, \eref{eq:bmom0} can be written as:
\begin{equation} \label{eq:bmom}
\rho \ddot{\ub} = \sum_{A,j,B} \bbb_{ikjl} u_{j,kl} \eb_i
\end{equation}
based on the push-forward of $\Bbb$ \cite[Eq. 4.2.34]{eringen2013linear}:
\begin{equation} \label{eq:smallbb}
\left[ \bbb \right]_{ijkl} = \frac{1}{\det(\Fb)} \sum_{J,L}  \left[ \Bbb \right]_{iJkL} \left[\Fb\right]_{jJ} \left[\Fb\right]_{lL} 
= \left[ \cbb \right]_{ijkl} + \left[ \sigmab \right]_{ik} \delta_{jl}
\end{equation}
where $\cbb$ is the push-forward of $\Cbb$ by the deformation gradient:
\begin{equation}
\left[ \cbb \right]_{ijkl} = \frac{1}{\det(\Fb)} \sum_{I,J,K,L}  \left[ \Cbb \right]_{ijkl} \left[\Fb\right]_{iI} \left[\Fb\right]_{jJ} \left[\Fb\right]_{kK} \left[\Fb\right]_{lL} 
\end{equation}

The Legendre-Hadamard criterion for dynamic stability requires that all infinitesimal plane waves 
\begin{equation}
\ub = \ab \cos(\kb\cdot\xb + \omega t) 
\end{equation}
have real-valued wave speeds.
Here, $a$ and $\pb$ are the amplitude and polarization of the displacement (such that $\ab = a \pb$ ), and $k$ and $\nb$ are the wave number and propagation direction (such that $\kb = k \nb$).
This leads to an eigenvalue problem for the dyad $\nb \otimes \pb$ and the strong ellipticity condition 
\begin{equation} \label{eq:strong_ellipicity}
(\nb \otimes \pb)^T : \bbb \, (\nb \otimes \pb) =
\left[ \bbb \right]_{ijkl} n_i p_j n_k p_l > 0
\end{equation}
This condition is satisfied when all the eigenvalues of the square matrix $\bbb_{(ij)(kl)}$ are all real and positive. 

The moduli that VASP and other codes calculate are derivatives of the current, Cauchy stress with respect to small strains about a given configuration, which is not $\cbb$.
To connect $\bbb$ to the moduli obtained from perturbing the system about a given (not necessarily stress-free reference) configuration, we start with the derivative of the Cauchy stress with respect to a displacement $\ub$ about a deformed state $\bar{\Fb}$
\begin{align} \label{eq:gateaux}
\left. \partial_\xb  \sigmab \right|_{\bar{\Fb}} \cdot \ub 
&= 
\left. \partial_\xb \left(  \frac{1}{\det(\Fb)} \Fb \Sb \Fb^T  \right) \right|_{\bar{\Fb}} \cdot \ub \nonumber \\
&= -\frac{1}{\det^2(\bar{\Fb})} \det \bar{\Fb} \left[ \tr \partial_\xb \ub  \right] \bar{\Fb} \bar{\Sb} \bar{\Fb}^T  
+ \frac{1}{\det(\Fb)}  \left[ \partial_\xb \ub \right] \bar{\Fb} \bar{\Sb} \bar{\Fb}^T \\
&+ \frac{1}{\det(\Fb)}                   \bar{\Fb} \, \partial_\Eb \bar{\Sb} \bar{\Fb}^T \, \frac{1}{2}  \left[ \partial_\xb^T \ub +  \partial_\xb^T \ub  \right]  \bar{\Fb}  \bar{\Fb}^T
+ \frac{1}{\det(\Fb)}                   \bar{\Fb} \bar{\Sb} \bar{\Fb}^T ( \partial_\xb^T \ub ) \nonumber
\end{align}
formed from the basic G\^ateaux derivatives in \cref{BonetWood}(Eqs. 3.69, 3.71, and 3.76).
We recognize that the third term on the right-hand side is $\cbb$ and $\bar{\sigmab}$ in the other terms, so that we can form the Fr\'echet derivative of the Cauchy stress with respect to the small strain measure $\epsilonb = \frac{1}{2}  \left( \partial_\xb \ub +  \partial_\xb^T \ub  \right) $ as:
\begin{equation} \label{eq:frechet}
\left[ \left. \partial_\epsilonb  \sigmab \right|_{\bar{\Fb}} \right]_{ijkl}
= - \bar{\sigmab}_{ij} \delta_{kl} 
+  \delta_{il} \bar{\sigmab}_{jk}  
+  \delta_{jk} \bar{\sigmab}_{il}  
+ \cbb_{ijkl}
\end{equation}
which is the typical moduli calculated by finite differences or perturbation in terms of the current stress $\bar{\sigmab}$ and push-forward of the tensor of the traditional elasticities $\Cbb$ to the current state.
\eref{eq:frechet} is identical in form to corresponding equations in the often cited \cref{wang1995mechanical}, and in the independently derived \cref{sinko2002ab}, but differs in the interpretation as moduli about a deformed state finitely far from the relaxed, stress-free material.

When \eref{eq:frechet} is combined with \eref{eq:smallbb}, the stability requirement \eqref{eq:strong_ellipicity} can be applied to:
\begin{equation}
\bbb_{ijkl} 
=  \delta_{kl} {\sigmab}_{ij}
-  \delta_{il} {\sigmab}_{jk}  
-  \delta_{jk} {\sigmab}_{il}  
+  \delta_{jl} {\sigmab}_{ik}  
+ \left[ \left. \partial_\epsilonb  \sigmab \right|_{\bar{\Fb}} \right]_{ijkl}
\end{equation}
For our purposes it suffices to find the stability conditions for an orthotropic modulus tensor $\left. \partial_\epsilonb  \sigmab \right|_{\bar{\Fb}}$ and a diagonal stress tensor $\sigmab = \sum_i \sigma_{ii} \eb_i \otimes \eb_i$.
Following \cref{mouhat2014necessary}, we obtain:
\begin{align}
\tilde{C}_{ii} >& 0, \ \  i \in {1,6}  \\
\tilde{C}_{ii} \tilde{C}_{jj} >& \tilde{C}^2_{ij}, \ \ i\neq j \in {1,3}  \\
\tilde{C}_{11} \tilde{C}_{22} \tilde{C}_{33} 
+ 2 \tilde{C}_{12} \tilde{C}_{23} \tilde{C}_{13} >&
\tilde{C}_{11} \tilde{C}^2_{23} +
\tilde{C}_{22} \tilde{C}^2_{13} +
\tilde{C}_{33} \tilde{C}^2_{12}
\end{align}
where 
$\tilde{C}_{ij} = {C}_{ij} + \frac{1}{2} \left( \sigma_{ii} + \sigma_{jj} \right)$, $i\neq j \in {1,3}$;
$\tilde{C}_{ij} = {C}_{ij} - \sigma_{kk}$, $i\neq j \in {4,6}$, $k-3 \neq i,j$;
and we have used $C_{ij}$ to denote the components of $\left[ \left. \partial_\epsilonb  \sigmab \right|_{\bar{\Fb}} \right]$ using traditional Voigt notation.
This reduces to 
\begin{equation}
C_{11} + 2 C_{12} + 3 p > 0, \ \ C_{11} - C_{12} > 0, \ \ C_{44} > 0
\end{equation}
for cubic symmetry and a hydrostatic pressure $\sigmab = -p \Ib$.
These stability criteria differ from those in \cref{wang1995mechanical} and  \cref{sinko2002ab} in that the shear conditions are unaffected by the pressure and the volumetric instability criterion on the bulk modulus $\frac{1}{3} \left( C_{11} + 2 C_{12} \right)$ is offset by the pressure only.

\end{document}